\documentclass[a4paper,11pt]{article}
\pdfoutput=1 

\usepackage{jcappub} 
                     
\usepackage{amsmath}
\usepackage{amsthm}
\usepackage{graphicx}
\usepackage{xspace}
\usepackage{color}
\usepackage{slashed}
\usepackage{units}
\usepackage{amssymb}

\def \L {\mathcal{L}} 

\def \vec#1{{\boldsymbol{#1}}}
\newcommand{\matrixx}[1]{\begin{pmatrix} #1 \end{pmatrix}} 
\newcommand{\hc}{\ensuremath{\text{h.c.}}}
\newcommand{\del}{\partial}
\newcommand{\dd}{\mathrm{d}}

\def \i {\mathrm{i}\mkern1mu} 

\allowdisplaybreaks

\newtheorem{theorem}{Theorem}
\numberwithin{theorem}{subsection}
\newtheorem{definition}[theorem]{Definition}

\newtheorem{lemma}[theorem]{Lemma}
\newtheorem{remark}{Remark}
\DeclareMathOperator{\diag}{diag}

\title{Cold light dark matter in extended seesaw models}

\author[a,b]{Sami Boulebnane,}

\author[a]{Julian Heeck,}

\author[a,b]{Anne Nguyen,}

\author[a]{Daniele Teresi}

\affiliation[a]{Service de Physique Th\'eorique, Universit\'e Libre de Bruxelles, Boulevard du Triomphe, CP225, 1050 Brussels, Belgium}

\affiliation[b]{Centre de Physique Th\'eorique, \'Ecole Polytechnique, Universit\'e Paris-Saclay, 91128 Palaiseau Cedex, France}

\emailAdd{Sami.Boulebnane@polytechnique.edu}

\emailAdd{Julian.Heeck@ulb.ac.be}

\emailAdd{Anne.Nguyen@polytechnique.edu}

\emailAdd{Daniele.Teresi@ulb.ac.be}

\abstract{We present a thorough discussion of light dark matter produced via freeze-in in two-body decays $A\to B$ DM. If $A$ and $B$ are quasi-degenerate, the dark matter particle has a cold spectrum even for keV masses. We show this explicitly by calculating the transfer function that encodes the impact on structure formation. As examples for this setup we study extended seesaw mechanisms with a spontaneously broken global $U(1)$ symmetry, such as the inverse seesaw. The keV-scale pseudo-Goldstone dark matter particle is then naturally produced cold by the decays of the quasi-degenerate right-handed neutrinos.}

\begin{document}

\maketitle

\flushbottom


\section{Introduction}

Cosmological and astrophysical evidence for dark matter (DM) requires an extension of the Standard Model (SM) of particle physics, which lacks a candidate that is sufficiently stable, massive, and dark. Considerable attention has been devoted to the search for DM particles with electroweak masses and couplings, so far without any undisputed evidence. An interesting alternative comes in the form of light DM, having in mind masses at the keV scale. Such DM particles are typically required to be produced non-thermally, implying small couplings to the SM~\cite{Hall:2009bx,Bernal:2017kxu}. They can nevertheless be searched for, especially if they are unstable and produce x-ray signals in their decay~\cite{Essig:2013goa}. A hint for a line-like signal with photon energy $\unit[3.5]{keV}$ was indeed observed recently in several astrophysical objects~\cite{Bulbul:2014sua,Boyarsky:2014jta,Boyarsky:2014ska,Iakubovskyi:2015dna,Cappelluti:2017ywp}; non-observation in other objects~\cite{Anderson:2014tza,Malyshev:2014xqa,Sekiya:2015jsa} makes the relevance of this signal difficult to assess~\cite{Adhikari:2016bei,Abazajian:2017tcc}. More data is required to settle this issue, but it provides an interesting jump-off point to speculate about its implications for new physics. Many possible models have been put forward, but arguably the simplest explanation of such an x-ray line would be the decay of a $\unit[7]{keV}$ DM scalar (fermion) into $\gamma\gamma$ ($\gamma \nu$).

We are not the first to point out that such keV DM particles could endanger the formation of small structures in our Universe, seeing as they have free-streaming lengths of order of Mpc if produced thermally. There are in fact some problems in structure-formation simulations that could be solved if DM would smear out smaller structures~\cite{Bode:2000gq,Adhikari:2016bei,Bullock:2017xww}, but they could also be artifacts of the non-inclusion of baryons~\cite{Fattahi:2016nld}. This issue is far from settled, and we have nothing to add to the discussion. Aside from $N$-body simulations, we however also have astrophysical observations of distant quasars at our disposal, which can be used to study small structures~\cite{Viel:2005qj}. Data from these Lyman-$\alpha$ spectral lines provides strong constraints on the free-streaming length of DM, typically quoted as a lower mass limit in the benchmark model of a thermal relic fermion.
Translating these limits to other scenarios requires precise knowledge of the DM momentum distribution function, which is itself determined by the DM production mechanism. In particular, many of the popular production mechanisms of a $\unit[7]{keV}$ DM particle are in tension with Lyman-$\alpha$ bounds~\cite{Merle:2014xpa,Schneider:2016uqi}, making necessary more involved ways to produce cold light DM if the x-ray hint is taken seriously~\cite{Merle:2013wta,Merle:2015oja,Konig:2016dzg,Heeck:2017xbu,Bae:2017tqn,Bae:2017dpt}.

As put forward recently by two of us (JH and DT), there exist several freeze-in production mechanisms that allow for light DM to be almost arbitrarily cold, and in particular easily accommodate a $\unit[7]{keV}$ mass without violating Lyman-$\alpha$ constraints~\cite{Heeck:2017xbu}. Our proposed scenarios require additional thermalized heavy particles which produce light DM either via scattering $A\,B\to C\,\mathrm{DM}$ or decays $A\to B\,\mathrm{DM}$.
In this article we will study the latter scenario in more detail, in particular providing the transfer functions necessary to assess Lyman-$\alpha$ bounds. We will also study extended seesaw models that naturally accommodate this production mechanism and in addition solve the neutrino mass problem of the SM. As discussed in detail in~\cite{Heeck:2017xbu}, the putative x-ray line can be generated in these models without spoiling the cold-enough production of DM.

The rest of this article is structured as follows: in Sec.~\ref{sec:freeze-in} we discuss the momentum distribution function of light DM produced by freeze-in decay, to then study its impact on structure formation by means of its transfer function.
In Sec.~\ref{sec:models} we delve into simple extended seesaw models that naturally lead to a light pseudo-Goldstone DM candidate coupled to quasi-degenerate heavy neutrinos. We conclude and summarize our work in Sec.~\ref{sec:conclusion}.
A number of appendices provide additional information for the interested reader. 
App.~\ref{sec:majorana_decays} lists relevant decay width formulae for Majorana fermions.
App.~\ref{sec:boltzmann_derivation} gives a derivation of the most general Boltzmann equation relevant for freeze-in production via two-body decays of thermalized particles. Finally, in App.~\ref{sec:perturbation_theory} we discuss the matrix perturbation theory for singular matrices such as the extended seesaw ones.

\section{Freeze-in of dark matter from decays}
\label{sec:freeze-in}

In this section we will discuss the momentum distribution function of light freeze-in DM produced via two-body decays of thermalized particles. This generalizes the analysis of Ref.~\cite{Heeck:2017xbu} for decays. 
We assume throughout that the new thermalized particles decay before Big Bang nucleosynthesis.

\subsection{Dark matter momentum distribution}
\label{sec:distribution}

The distribution function $f(p,T)$ of DM from the freeze-in decay $A\to B\,\mathrm{DM}$ is conveniently rewritten in terms of the dimensionless parameters $x=|\vec{p}|/T$ and $r = m_A/T$. A full derivation is given in App.~\ref{sec:boltzmann_derivation}, here we only show the result using a Maxwell--Boltzmann distribution for the thermalized mother particle $A$ (with $g_A$ internal degrees of freedom):
\begin{align}
\frac{\partial f(x,r)}{\partial r}  = \frac{g_A\,S\,\Gamma\,M_0\, r^2\, \sinh\left[\frac{m_A p_\mathrm{DM} x}{m_\mathrm{DM}^2}\right]}{p_\mathrm{DM} x \sqrt{m_A^2 x^2 + m_\mathrm{DM}^2 r^2}} \exp\left[- \frac{m_A^2-m_B^2+m_\mathrm{DM}^2}{2 m_A m_\mathrm{DM}^2} \sqrt{m_A^2 x^2 + m_\mathrm{DM}^2 r^2}\right] .
\label{eq:fullMBdistribution}
\end{align}
Here, $\Gamma$ is the partial decay width of $A\to B\,\mathrm{DM}$ in the rest frame of $A$ and
\begin{align}
p_\mathrm{DM} = \frac{\sqrt{(m_A^2-(m_B+m_\mathrm{DM})^2)(m_A^2-(m_B-m_\mathrm{DM})^2)}}{2 m_A} 
\label{eq:pDM}
\end{align}
the total three-momentum of the DM particle in that frame.
$S$ is a symmetry factor that is equal to 2 if $B = \mathrm{DM}$ and 1 otherwise; $M_0 \equiv M_\text{Pl} \sqrt{45/(4\pi^3 g_*)}$ is the rescaled Planck mass.

The DM distribution $f(x,R)$ at ``time'' $R$ can be obtained from Eq.~\eqref{eq:fullMBdistribution} by integrating over $r$ from $0$ to $R$, which is in general not possible analytically, but straightforward numerically. The final DM abundance today ($R\to \infty$) then follows as a further integral over all momenta, normalized to the critical density~\cite{Kolb:1990vq,Heeck:2017xbu},
\begin{align}
\Omega_\text{DM}h^2 = \frac{s_0 m_\mathrm{DM}}{\rho_\mathrm{crit}/h^2}\left[\frac{45/(4\pi^4)}{g_*(T_\mathrm{prod})} \int_0^\infty \dd x \, x^2 f(x,\infty) \right].
\label{eq:abundance}
\end{align}
Similarly, the mean DM momentum at production time $r_\mathrm{prod}$ can be obtained from
\begin{align}
 \left\langle \frac{p}{T}\right\rangle_\mathrm{prod} =\langle x \rangle = \frac{\int \! \dd^3  \vec p \, |\vec p|  f(\vec p,r_{\rm prod})}{T_\mathrm{prod}\int \! \dd^3  \vec p \, f(\vec p,r_{\rm prod})} = \frac{\int_0^\infty \dd x \, x^3 f(x,r_\mathrm{prod})}{\int_0^\infty \dd x \, x^2 f(x,r_\mathrm{prod})} \,.
\label{eq:mean_pT}
\end{align}
Since our derivation assumes a constant $g_*$, we will let $r_\mathrm{prod}\to\infty$ in order to obtain the mean momentum \emph{today}, taking care of the entropy dilution in the SM bath by hand~\cite{Petraki:2007gq}:
	\begin{align}
	\left\langle \frac{p}{T}\right\rangle = \left(\frac{g_*(T_0)}{g_*(T_\mathrm{prod})}\right)^{1/3}\left\langle \frac{p}{T}\right\rangle_\mathrm{prod}  \simeq 0.32 \left(\frac{106.75}{g_*(T_\mathrm{prod})}\right)^{1/3} \left\langle \frac{p}{T}\right\rangle_\mathrm{prod} .
	\label{eq:poverTtoday}
	\end{align}
The production temperature $T_\mathrm{prod}$ is around $T_\mathrm{prod}/m_A \sim \frac13$--$\frac15$~\cite{Hall:2009bx,Frigerio:2011in}, which we will assume to be above the electroweak scale for the most part. Since the mother particle $A$ (and potentially $B$) is in equilibrium with the SM by assumption, it will contribute to $g_*$, but the effect is mild unless many new particles are introduced.
If more than one DM production process exists, e.g.~several decay channels $A_i\to B_j\,\text{DM}$, their effect on $f$ is simply additive because we are in the freeze-in regime~\cite{McDonald:2001vt,Hall:2009bx,Bernal:2017kxu} where the backreaction of $f$ on the thermal plasma is negligible.

We are interested in the limit of very light DM, and in particular $m_\mathrm{DM}\ll m_A-m_B$. Taking the limit $m_\mathrm{DM}\to 0$ in Eq.~\eqref{eq:fullMBdistribution} allows us to perform the integration over $r$, resulting in a simple DM distribution function today~\cite{Heeck:2017xbu},
\begin{align}
f(x,\infty) = \frac{2 \sqrt{\pi } g_A\,S\,\Gamma\,M_0}{m_A^2 \Delta ^3} \sqrt{\frac{\Delta}{x}} \exp\left(-\frac{x}{\Delta}\right) , \quad \text{ with } \quad
\Delta \equiv 1-\frac{m_B^2}{m_A^2}\,,
\label{eq:approxMBdistribution}
\end{align}
and equally simple expressions for DM abundance and mean momentum,
\begin{align}
\Omega_\text{DM}h^2 &= \frac{135}{8\pi^3 g_*(T_\mathrm{prod})} \frac{s_0 m_\mathrm{DM}}{\rho_\mathrm{crit}/h^2}\frac{g_A S\, \Gamma\,M_0}{m_A^2} \simeq 10^{24}\,\frac{g_A S\, \Gamma\,m_\mathrm{DM}}{m_A^2} \,, &&
\left\langle \frac{p}{T}\right\rangle_\mathrm{prod} =  \frac{5}{2} \,\Delta \,.
\label{eq:massless_approx}
\end{align}
In particular, the mean DM momentum becomes smaller the more degenerate $A$ and $B$ are, which implies that DM becomes colder~\cite{Heeck:2017xbu}. Before we verify this statement by looking at structure formation, let us first argue that it is sufficient to work with the analytical approximation of Eq.~\eqref{eq:approxMBdistribution}.

\begin{figure}
\centering
\includegraphics[width=0.9\textwidth]{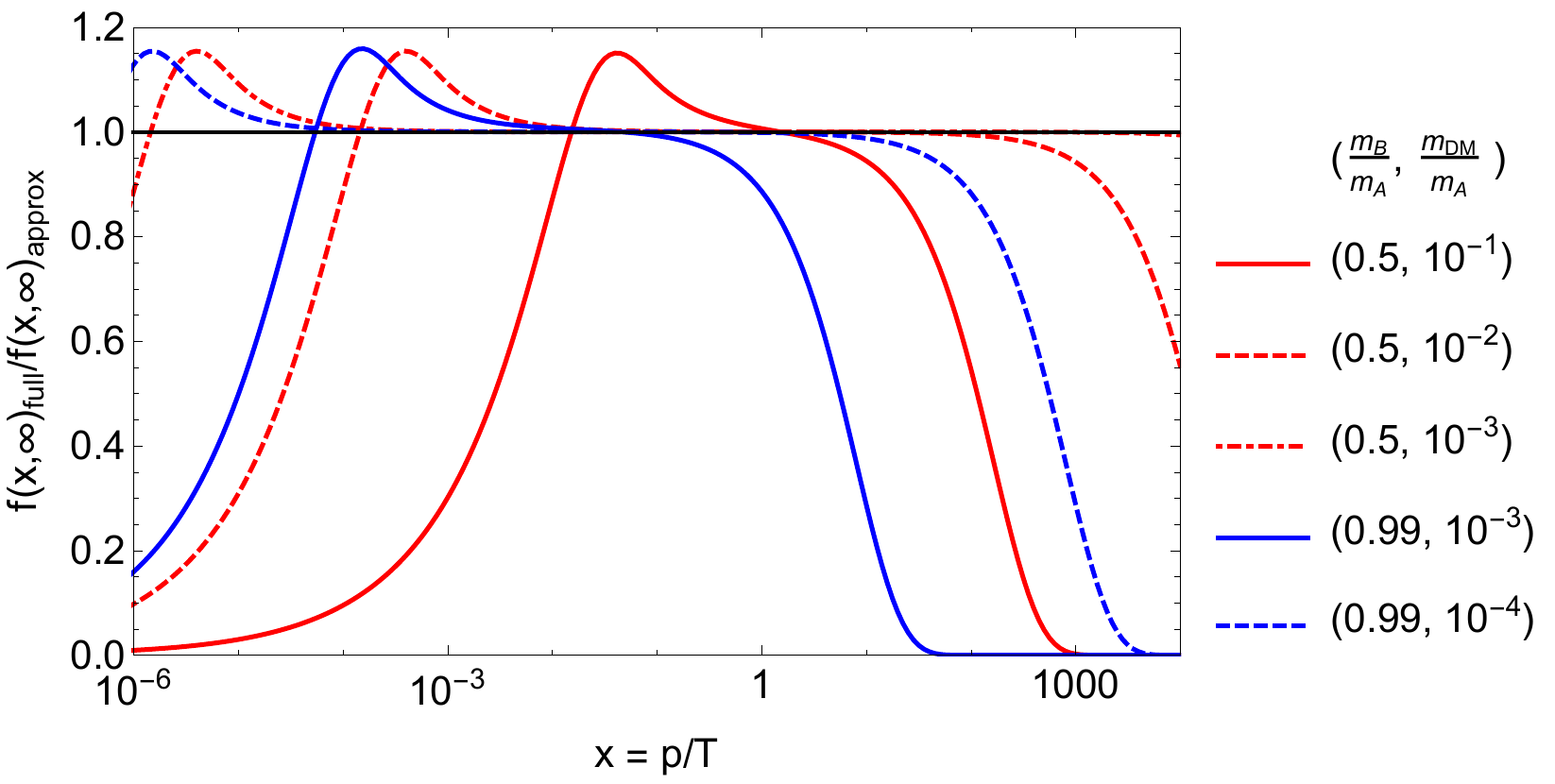}
\caption{
Ratio of DM distribution with full DM mass dependence (Eq.~\eqref{eq:fullMBdistribution}) over approximate $m_\text{DM} = 0$ result (Eq.~\eqref{eq:approxMBdistribution}) for different values of the masses in the decay $A \to B \, \mathrm{DM}$.
}
\label{fig:mass_effects_on_f}
\end{figure}

\begin{figure}
\centering
\includegraphics[width=0.6\textwidth]{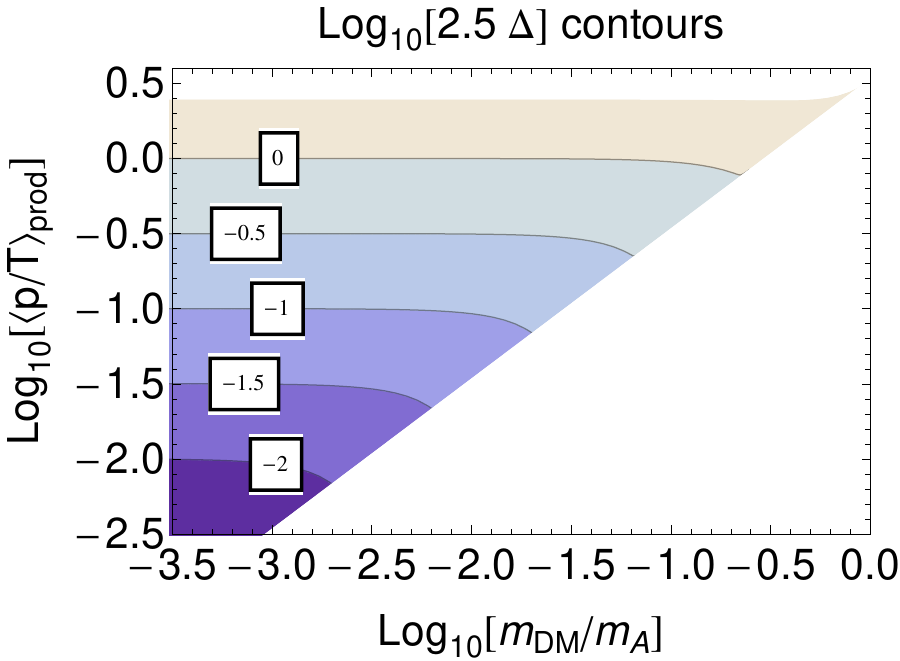}
\caption{
Average $\langle x\rangle =\left\langle p/T\right\rangle_\mathrm{prod}$ as a function of the DM mass in units of $m_A$. The contours correspond to fixed values of $\tfrac52\Delta = \tfrac52 \left(1- \tfrac{m_B^2}{m_A^2}\right)$, which matches $\langle x\rangle$ for $m_\mathrm{DM}\ll m_A-m_B$ (see Eq.~\eqref{eq:massless_approx}).
In the white region the decay $A\to B\,\mathrm{DM}$ is kinematically forbidden.
}
\label{fig:avg_x_vs_DM_mass}
\end{figure}

In Fig.~\ref{fig:mass_effects_on_f} we show the ratio of the full distribution function over the analytical approximation of Eq.~\eqref{eq:approxMBdistribution} for various values of $m_B$ and $m_\mathrm{DM}$. As can be seen, keeping the DM mass nonzero cuts off the distribution function both for low and high $x$. This is not particularly important in practice, since these extremal values are anyway suppressed in the relevant function $x^2 f(x,\infty)$. Interestingly, the analytical approximation typically matches the numerical result  in the region of highest probability $x \sim 2.5 \, \Delta$.  Furthermore, one can see that the full solution will actually lead to a smaller mean momentum, making DM colder still (Fig.~\ref{fig:avg_x_vs_DM_mass}). However, the effect is negligible unless the hierarchy $m_\mathrm{DM}\ll m_A-m_B$ is not realized.
Notice that the limit $m_\mathrm{DM}\to m_A-m_B$ leads to a vanishing DM momentum in the \emph{rest frame} of $A$ (Eq.~\eqref{eq:pDM}), but not in the thermal bath frame, where $A$ is thermally distributed~\cite{Heeck:2017xbu}. Numerically, we find the finite average DM momentum $\langle x\rangle \simeq 3.4 \,m_\mathrm{DM}/m_A \simeq 3.4 (1-m_B/m_A)$ in this limit. Obviously the DM production becomes arbitrarily inefficient for $m_\mathrm{DM}\to m_A-m_B$, making this limit rather uninteresting. Therefore, as expected, the DM mass effects are negligible except for the region of \emph{total} phase-space closure of the decay $A\to B\,\mathrm{DM}$, $m_\mathrm{DM}\simeq m_A-m_B$, in which case the DM momentum is even further suppressed. We will therefore use the massless DM approximation of Eq.~\eqref{eq:approxMBdistribution} for the distribution function in the following. 

\subsection{Structure formation}
\label{sec:structure}

Thermalized DM particles in the keV range are usually considered dangerous for structure formation due to their large free-streaming length, which implies a wash-out of small scale structures. A popular probe for this comes from the Lyman-$\alpha$ forest, i.e.~light from distant quasars~\cite{Viel:2005qj}. These are typically stronger than other limits, for instance from satellite counting. Limits are usually derived for the benchmark model of a thermal relic fermion with mass $m_\mathrm{TR}$, modeled after the known neutrinos. If such a particle makes up all of the DM abundance, current limits range from $m_\mathrm{TR}\gtrsim \unit[4.09]{keV}$~\cite{Baur:2015jsy}, $\unit[4.65]{keV}$~\cite{Yeche:2017upn,Baur:2017stq} to $\unit[5.3]{keV}$~\cite{Irsic:2017ixq}, depending on the combination of data sets and the peculiarities of the analysis. In this article we will use $\unit[4.65]{keV}$ as a fairly conservative limit, but our results can be easily rescaled.

As emphasized by many authors before us, structure formation mass limits depend strongly on the DM momentum distribution function, which in turn depends on the DM production mechanism. 
A naive way of translating the thermal relic limits on $m_\mathrm{TR}$ to other models (with DM mass $m_\mathrm{DM}$) is to set equal their free-streaming lengths~\cite{Colombi:1995ze}, which effectively depends on the mean DM momentum $\langle p/T\rangle$. This leads to the simple formula~\cite{Shaposhnikov:2006xi,Bezrukov:2014nza,Adhikari:2016bei,Heeck:2017xbu}
\begin{align}
\unit[5.1]{keV}\left(\frac{106.75}{g_*(T_\mathrm{prod})}\right)^{\frac{1}{3}} \left(\frac{m_\mathrm{TR}}{\unit[4.65]{keV}}\right)^{\frac{4}{3}} \left\langle \frac{p}{T}\right\rangle_\mathrm{prod} \lesssim m_\mathrm{DM}  .
\label{eq:translation2}
\end{align}
Using our result from Eq.~\eqref{eq:massless_approx} and assuming DM production above the electroweak scale, this implies $\unit[12.8]{keV} \lesssim m_\mathrm{DM}/\Delta$. Since $\Delta$ can be made arbitrarily small, the DM mass can in principle be lowered even far below the keV scale.
If DM is a fermion, there still exists a lower mass bound of order hundreds of eV that holds independently of the production mechanism simply due to Fermi--Dirac statistics~\cite{Tremaine:1979we} (for an updated analysis using various assumptions see, e.g.,~\cite{Boyarsky:2008ju}); for bosonic DM on the other hand, we can apparently make DM arbitrarily light if $\Delta$ is tiny.
We stress that this is qualitatively different from the so-called misalignment mechanism, popularized through axion DM~\cite{Olive:2016xmw}, which uses \emph{classical field oscillations} to obtain cold DM~\cite{Marsh:2015xka}. 

Independent of the production mechanism, one can still obtain a lower mass limit around $\unit[10^{-21}]{eV}$~\cite{Irsic:2017yje,Armengaud:2017nkf,Kobayashi:2017jcf} for bosonic DM from structure formation and Lyman-$\alpha$ due to the macroscopic de~Broglie wavelength of such a light particle. This \emph{fuzzy DM} scenario has recently been popularized as an alternative to warm DM in solving small-scale structure issues~\cite{Hu:2000ke,Hui:2016ltb}.
With our production mechanism it is not possible to reach the fuzzy DM regime, at least not with our approximation of TeV-scale $A$; this is because we cannot compensate an arbitrarily small $m_\mathrm{DM}$ in $\Omega_\mathrm{DM}h^2$ (Eq.~\eqref{eq:massless_approx}) by a larger width $\Gamma$. Demanding $\Gamma (A\to B\,\mathrm{DM})< m_A$ gives $m_A < 10^{25} m_\mathrm{DM}$, so fuzzy DM values unavoidably require rather light $A$ and $B$ particles, which introduces many additional constraints.
Demanding for simplicity $m_A$ to be heavier than TeV to satisfy all our assumptions then gives $\unit[10^{-13}]{eV}$ as the lowest DM mass achievable with our production mechanism. This is already very optimistic, seeing as $\Gamma\simeq m_A$ would imply large couplings that make possible other DM production mechanisms, e.g.~via scattering, as we are going to show explicitly below. 

More importantly, DM with mass below keV requires a large DM \emph{number density} in order to achieve $\Omega_\text{DM}h^2\simeq 0.1$. This is in conflict with our freeze-in approximation, i.e.~that the DM abundance is negligible and we can ignore inverse reactions such as $B\,\mathrm{DM} \to A$. 
To estimate the region of freeze-in validity, we calculate the thermally-averaged reaction rate for the inverse process $B\,\mathrm{DM} \to A$ relative to $A\to B\,\mathrm{DM}$, assuming that the DM distribution is still given by the freeze-in formula of Eq.~\eqref{eq:fullMBdistribution}. This ratio peaks at $T\simeq m_A/5$, so we have to demand conservatively	
\begin{align}
	\left.\frac{\int \dd \text{PS}\,f_B f_\text{DM} |\mathcal{M} (B \,\text{DM} \to A)|^2 }{\int \dd \text{PS}\, f_A |\mathcal{M} (A \to B \,\text{DM})|^2 }\right|_{T\simeq m_A/5} \simeq 0.18\, \frac{g_A\Gamma  (A \to B \,\text{DM}) M_0}{\Delta^3 m_A^2} \stackrel{!}{<} 1
\end{align}
for freeze in. Here, $\dd \text{PS}$ denotes the appropriate phase-space integration measure including energy--momentum conservation and $\mathcal{M}$ the matrix element, which is the same for both directions. Using Eq.~\eqref{eq:massless_approx} to translate the partial width $\Gamma  (A \to B \,\text{DM})$ into the DM abundance, we find the inequality $\unit[10]{eV}/\Delta^3 <m_\text{DM}$ for freeze-in DM. Below this value, the DM distribution will start to differ from Eq.~\eqref{eq:fullMBdistribution} and require a full solution of the integro-differential Boltzmann equations, with is beyond the scope of this article. In combination with the Lyman-$\alpha$ bound, this implies that our calculations are trustworthy down to DM masses of $\sim \unit[2]{keV}$. 
Note that both rates $B\,\mathrm{DM} \leftrightarrow A$ can still be smaller than the Hubble rate when the above inequality is violated, so DM is not automatically in equilibrium with the SM. The shape of the DM distribution will however move towards a thermal one when $B\,\mathrm{DM} \to A$ becomes relevant.

We have yet to verify our translation formula from above.
Eq.~\eqref{eq:translation2} relies on the free-streaming length as a good measure of structure wash out. Implicitly, it assumes that the DM momentum distribution possesses a relevant mean value that characterizes it. While this might be valid for the distribution of Eq.~\eqref{eq:approxMBdistribution} we use as an approximation, it is obviously not valid in general, since the distribution function can in principle be arbitrarily complicated and without useful mean~\cite{Murgia:2017lwo}.
Even sticking to our freeze-in production via decays, as soon as several channels $A_i\to B_j \mathrm{DM}$ with different $\Delta_{ij}$ and branching ratios contribute, $f(x,r)$ can have a complicated structure with several peaks~\cite{Heeck:2017xbu}. It is in those situations that it is necessary to go beyond the free-streaming-length approximation and delve into the actual calculation of structure formation~\cite{Colombi:1995ze,Bezrukov:2014qda,Konig:2016dzg,Murgia:2017lwo}.

\begin{figure}
\centering
\includegraphics[width=0.8\textwidth]{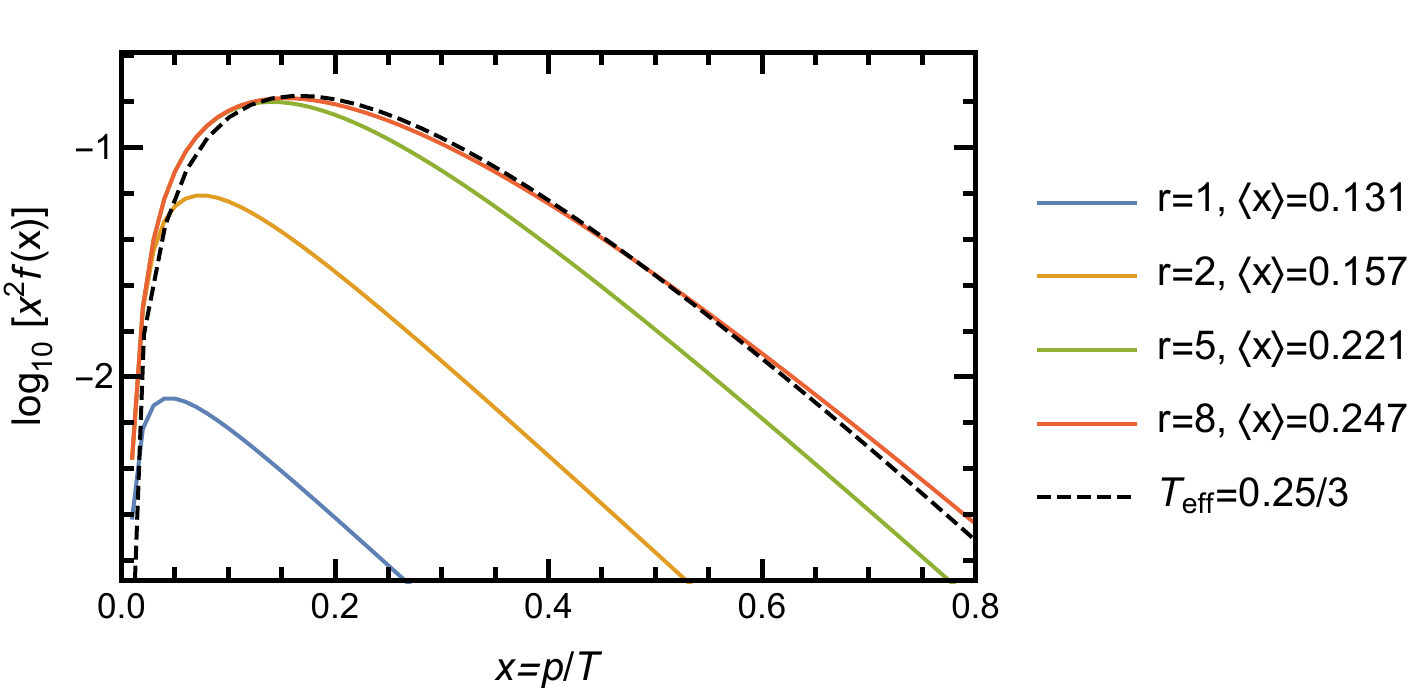}
\caption{
The distribution function for the decay $A \to B\,\mathrm{DM}$ with $\Delta = 0.1$ (for several values of the time variable $r$) compared to a rescaled thermal one, with effective temperature given by $T_{\rm eff} = \langle x \rangle/3$.
}
\label{fig:distribution}
\end{figure}

\begin{figure}
\centering
\includegraphics[width=0.7\textwidth]{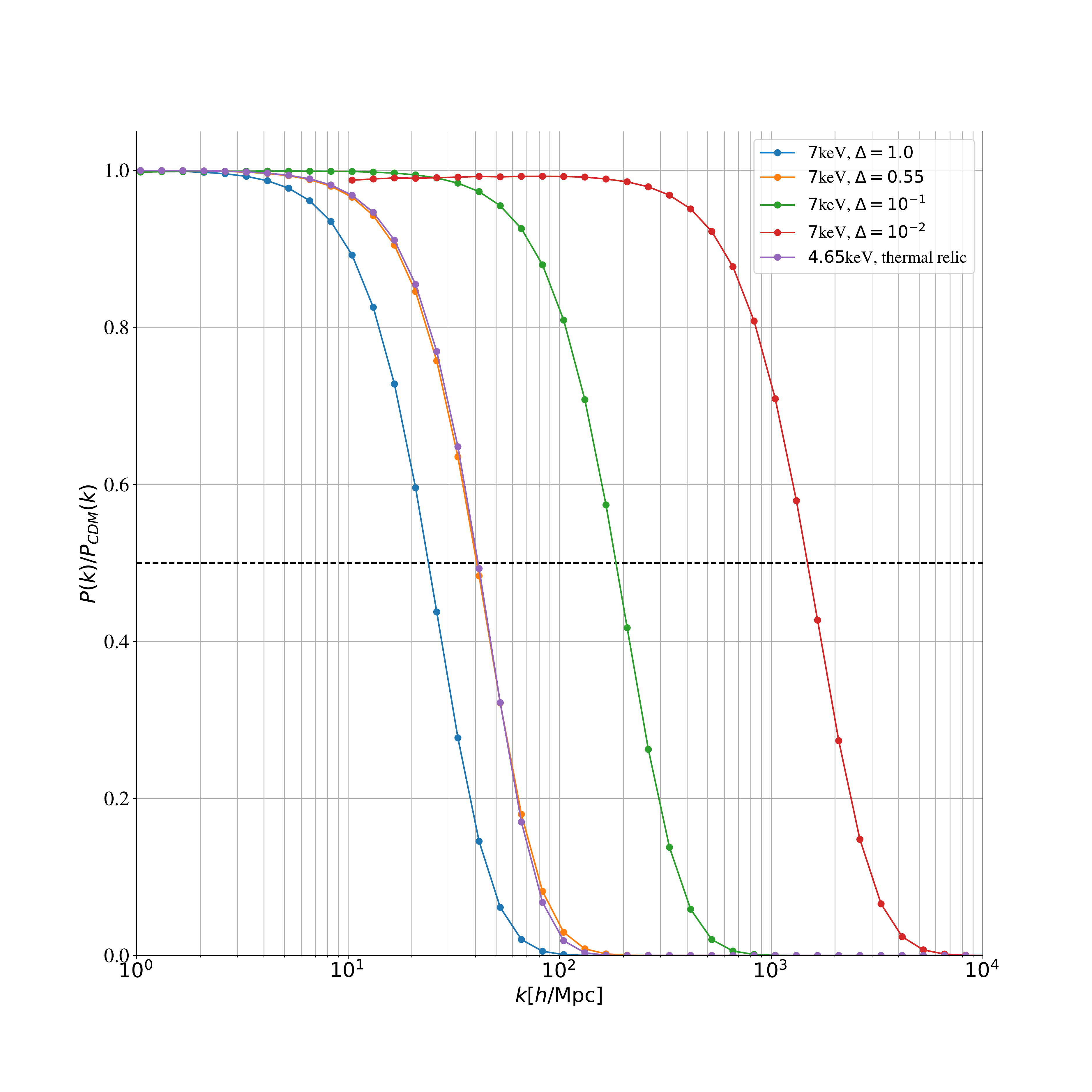}
\caption{
Transfer functions for DM produced via $A\to B\,\text{DM}$ with different $\Delta = 1-m_B^2/m_A^2$ as well as the transfer function of a thermal relic with mass $\unit[4.65]{keV}$ (purple), corresponding to the current Lyman-$\alpha$ limit~\cite{Yeche:2017upn,Baur:2017stq}.
}
\label{fig:4650eV_thermal_relic_7keV_many}
\end{figure}

It is beyond our scope to perform $N$-body simulations to study the impact of our models on structure formation. Instead, we implemented our distribution function of Eq.~\eqref{eq:approxMBdistribution} in the Boltzmann solver \texttt{CLASS}~\cite{Blas:2011rf,Lesgourgues:2011rh} (Cosmic Linear Anisotropy Solving System) to obtain the transfer function $\mathcal{T}(k)$ for a given wavenumber $k$,
\begin{align}
\mathcal{T}^2(k) \equiv \frac{P(k)}{P_\mathrm{CDM}(k)}\,,
\end{align}
$P$ and $P_\mathrm{CDM}$ being the power spectra for our DM model and cold DM, respectively.
The necessary cosmological parameters have been taken from Planck, specifically the dataset combination ``Planck 2015 TT, TE, EE+lowP''~\cite{Ade:2015xua}.
This transfer function is then compared to the function $\mathcal{T}_\mathrm{TR}$ we obtain for a thermal relic of mass $m_\mathrm{TR} = \unit[4.65]{keV}$, which we take as an exclusion region. Following Ref.~\cite{Konig:2016dzg}, we regard a model as excluded if $\mathcal{T}^2(k) \leq \mathcal{T}^2_\mathrm{TR}(k)$ for all $k$ smaller than the half-mode $k_{1/2}$, defined via $\mathcal{T}^2(k_{1/2}) = 1/2$.
This procedure is very robust in our case, since the distribution function is very close to a rescaled thermal one (see Fig.~\ref{fig:distribution}) and, consequently, the shape of our transfer function is almost identical to the thermal relic one (see Fig.~\ref{fig:4650eV_thermal_relic_7keV_many}).  In particular, a thermal relic of mass \unit[4.65]{keV} gives the same transfer function as our distribution function with $\Delta=0.55$ and $m_\mathrm{DM} = \unit[7]{keV}$, in accordance with Eq.~\eqref{eq:translation2}.
For $\Delta$ between $1$ and $10^{-2}$, we obtain an approximate expression for the half-mode $k_{1/2}$ of the form
\begin{align}
k_{1/2} \simeq \unit[25]{\frac{h}{Mpc}} \left( \frac{m_\mathrm{DM}}{\unit[7]{keV}\, \Delta} \right)^{0.9} \stackrel{!}{>} k_{1/2}^\mathrm{TR} \simeq \unit[41]{\frac{h}{Mpc}} ,
\label{eq:halfmode_single_decay}
\end{align}
which slightly improves on Eq.~\eqref{eq:translation2}, with 5\% discrepancy. The difference turns out to be minor, which shows that the mean-momentum and free-streaming length are useful quantities in the single-decay scenario.

If multiple decay channels are open, the DM distribution function is simply the sum of the different channels. For simplicity we will still consider only one mother particle $A$, but with different decay channels $A\to B_j \,\mathrm{DM}$ characterized by $\Delta_j = 1- m_{B_j}^2/m_A^2$ and branching ratios $\mathrm{BR}_j$. The function $x^2 f(x,\infty)$ then has a multi-peak form with moments
\begin{align}
\langle x\rangle = \frac{5}{2} \frac{\sum_j \mathrm{BR}_j \Delta_j}{\sum_j \mathrm{BR}_j }\,, &&
\langle x^2\rangle = \frac{35}{4} \frac{\sum_j \mathrm{BR}_j \Delta_j^2}{\sum_j \mathrm{BR}_j }\,, &&
\langle x^z\rangle = \frac{4\,\Gamma\left(z+\frac{5}{2}\right)}{3\sqrt{\pi}} \frac{\sum_j \mathrm{BR}_j \Delta_j^z}{\sum_j \mathrm{BR}_j }\,,
\end{align} 
the last equation with the Gamma function being valid for $\mathrm{Re}(z)\geq -5/2$, far more general than what is needed here.
We can also define the standard deviation $\sigma \equiv \sqrt{\langle x^2\rangle-\langle x\rangle^2}$.
If one decay channel dominates, it is easy to verify that $\sigma/\langle x\rangle  = \sqrt{2/5}<1$, so the mean momentum is a useful quantity, as explicitly verified above. In the presence of several decays, on the other hand, one can have $\sigma > \langle x\rangle$, making the mean $\langle x\rangle$ less useful~\cite{Merle:2015oja}. This happens essentially when channels exist which have the same $\mathrm{BR}_j \Delta_j$ but vastly different $\Delta_j$.
In this case one indeed finds that Eq.~\eqref{eq:halfmode_single_decay}, with $\Delta \to \tfrac{2}{5} \langle x\rangle$, is a bad estimator for $k_{1/2}$. Remarkably, the \emph{shape} of $\mathcal{T}^2(k)$ remains the same independent of the number of decay channels, only the position $k_{1/2}$ changes. This implies that $\mathcal{T}^2(k)$ or $k_{1/2}$ only depends on \emph{one} quantity to be build from $\{\mathrm{BR}_j, \Delta_j\}$. Replacing $\Delta \to \Delta_\text{eff}$ in Eq.~\eqref{eq:halfmode_single_decay} with
\begin{align}
\Delta_\mathrm{eff} = \left( \frac{\sum_j \mathrm{BR}_j \Delta_j^\eta}{\sum_j \mathrm{BR}_j } \right)^{1/\eta	} ,
\label{eq:Delta_eff}
\end{align}
turns out to be an excellent ansatz to generalize Eq.~\eqref{eq:halfmode_single_decay}, with $\eta = 1.9$ obtained from a fit. We have verified this formula for many $\{\mathrm{BR}_j, \Delta_j\}$ points with up to three decays and found a percent-level agreement.

To sum up our results for this part, we have seen that for light DM produced from several decays $A\to B_j \text{DM}$ with $\Delta_j = 1- m_{B_j}^2/m_A^2$ and $m_\text{DM}\ll m_A-m_{B_j}$, the Lyman-$\alpha$ forest sets the constraint
\begin{align}
k_{1/2} \simeq \unit[25]{\frac{h}{Mpc}} \left( \frac{m_\mathrm{DM}}{\unit[7]{keV}\, \Delta_\mathrm{eff}} \right)^{0.9} \stackrel{!}{>} k_{1/2}^\mathrm{TR} \simeq \unit[41]{\frac{h}{Mpc}} ,
\end{align}
with $\Delta_\text{eff}$ from Eq.~\eqref{eq:Delta_eff} with $\eta = 1.9$.
In other words, in our model of DM genesis via the decays of a massive particle in equilibrium, we can go down to DM masses
\begin{align}
m_\text{DM} > \unit[12]{keV} \left(\frac{\sum_j \mathrm{BR}_j \Delta_j^\eta}{\sum_j \mathrm{BR}_j}\right)^{1/\eta} \,, && \text{ with } && \eta \simeq 1.9\,,
\label{eq:final_Ly_alpha_limit}
\end{align}
without violating the Lyman-$\alpha$ bound that corresponds to a $\unit[4.65]{keV}$ thermal relic.
Eq.~\eqref{eq:final_Ly_alpha_limit} is the appropriate generalization of Eq.~\eqref{eq:halfmode_single_decay} in the presence of several decay channels.
In the following we will explore particle-physics models that lead to a small $\sum_j \mathrm{BR}_j \Delta_j^\eta$ in order to allow for DM masses at the keV scale.

\section{Extended seesaw models}
\label{sec:models}

In Sec.~\ref{sec:freeze-in} we have discussed how light DM produced by freeze-in decays $A\to B\,\mathrm{DM}$ affects structure formation and Lyman-$\alpha$ limits. We have established that bosonic DM can have keV masses without violating Lyman-$\alpha$ constraints, as long as $A$ and $B$ are somewhat degenerate and much heavier than the DM particle. In the second part of this article we will discuss simple particle-physics models that realize this scenario.
The obvious choice for a light bosonic DM particle is a pseudo-Goldstone boson of some global symmetry, as this can ensure a small mass without fine-tuning~\cite{Frigerio:2011in}. Plenty of candidates have been discussed in the literature already, be it majorons~\cite{Rothstein:1992rh,Berezinsky:1993fm,Lattanzi:2007ux,Bazzocchi:2008fh,Frigerio:2011in,Lattanzi:2013uza,Queiroz:2014yna,Garcia-Cely:2017oco}, connected to the lepton symmetry $U(1)_L$~\cite{Chikashige:1980ui,Schechter:1981cv}, familons~\cite{Frigerio:2011in}, connected to family symmetries~\cite{Wilczek:1982rv,Reiss:1982sq}, and axions (or axion-like) particles~\cite{Higaki:2014zua,Jaeckel:2014qea}, connected to the Peccei--Quinn symmetry $U(1)_\text{PQ}$~\cite{Kim:1986ax}.
Our examples below will be modeled after majorons in order to simplify the discussion. Note that we will not concern ourselves with the origin of the DM mass, but simply assume an explicit breaking term in the scalar potential. In all the cases below, the radiative decay of DM into photons can be generated via mixing with either the SM scalar, $Z$ boson or anomalous couplings to photons. The important point is that the couplings involved in the radiative decay do not spoil the successful production of cold DM~\cite{Heeck:2017xbu}. 

In the standard singlet majoron model~\cite{Chikashige:1980ui,Schechter:1981cv,Garcia-Cely:2017oco,Heeck:2017xbu}, the decays $N_i\to N_j J$ among the heavy neutrino states are suppressed compared to the decays into the light neutrinos $N_i \to \nu_j J$. The dominant part of the majoron DM is then produced with $\langle x\rangle = 5/2$, too warm for our purposes.
A solution was already put forward in Ref.~\cite{Heeck:2017xbu} in the form of extended seesaw mechanisms by assigning different $U(1)'$ charges to some right-handed neutrinos, thereby inducing faster $N_i \to N_j J$ decays. 

We are interested in minimal models, extending the SM only by gauge singlet fields for simplicity. To keep the number of parameters small, we also assume only one complex scalar field $\sigma$ to break the $U(1)'$ symmetry, with $J = \sqrt{2}\, \mathrm{Im}(\sigma) $ being the Goldstone boson of interest for DM.\footnote{One can also aim at generating all mass entries spontaneously, which requires additional scalars~\cite{Rojas:2017sih,Humbert:2015epa}.} In this setup, we have to introduce some right-handed neutrinos $N_R$ that carry the same $U(1)'$ charge as the SM neutrinos $\nu_L$ in order to generate neutrino masses. To get a different phenomenology from the singlet majoron scenario, we further introduce a number of right-handed fermions $S_R$ with different $U(1)'$ charge.
Playing with the $U(1)'$ charges, one can identify three interesting cases:
\begin{enumerate}
	\item \emph{Inverse Seesaw (IS)}: assigning $X(\nu_L) = X(N_R) =-X(\sigma)= 1$ and $X(S_R)=0$, the neutral fermion mass matrix in the basis $(\nu_L, N_R^c, S_R^c)$ takes the form
	\begin{align}
	\mathcal{M}_\text{IS} = \matrixx{0 & m_D & 0 \\ m_D^T & 0 & M \\ 0 & M^T & \mu} , \quad \text{ with } \quad M = \lambda \langle \sigma \rangle\,.
	\label{eq:ISmassmatrix}
	\end{align}
	For $\mu, m_D \ll M$, this scenario has been dubbed the inverse seesaw~\cite{Mohapatra:1986aw,Mohapatra:1986bd,GonzalezGarcia:1988rw,Frere:1989xb}.
	One typically sets $\# \nu_L = \# N_R = \# S_R = 3$, but even $\# N_R = \# S_R = 2$ is viable~\cite{Abada:2014vea}. 
	
	\item \emph{Extended Inverse Seesaw (EIS)}: assigning $X(\nu_L) = X(N_R) = 1$, $X(\sigma)=2$, and $X(S_R)=-1$, we find
	\begin{align}
	\mathcal{M}_\text{EIS} = \matrixx{0 & m_D & 0 \\ m_D^T & \mu_1 & M \\ 0 & M^T & \mu_2} , \quad \text{ with } \quad \mu_j = \lambda_j \langle \sigma \rangle\,.
	\label{eq:EISmassmatrix}
	\end{align}
	For $\mu_j, m_D \ll M$, this is an extended inverse seesaw mechanism and works with the same number of states. The majoron of this model with $\mu_1=0$ was already discussed in Ref.~\cite{GonzalezGarcia:1988rw}. 

	\item \emph{Extended Seesaw (ES)}: assigning $X(\nu_L) = X(N_R) = 0$, $X(\sigma)= X(S_R)=1$, we find
	\begin{align}
	\mathcal{M}_\text{ES} = \matrixx{0 & m_D & 0 \\ m_D^T & \mu & M \\ 0 & M^T & 0} , \quad \text{ with } \quad M = \lambda \langle \sigma \rangle\,.
	\label{eq:ESmassmatrix}
	\end{align}
	Although at first sight just a special case of the EIS, the above is not an \emph{inverse} seesaw mechanism at tree level. The case $\# \nu_L = \# N_R=3$, $\# S_R =1$ has been dubbed minimal extended seesaw~\cite{Barry:2011wb,Zhang:2011vh,Heeck:2012bz},\footnote{The main motivation to study $\mathcal{M}_\text{ES}$ in the past was the occurrence of light sterile neutrinos in the limit $m_D \lesssim M \ll \mu$~\cite{Ma:1995gf,Chun:1995js,Barry:2011wb,Zhang:2011vh,Heeck:2012bz,Adhikari:2016bei}.}and so we will refer to the more general scenario as Extended Seesaw. The number of massless states obtained by diagonalizing $\mathcal{M}_\text{ES}$ is $\# \nu_L +\# S_R- \# N_R$.
	Even though the active neutrinos do not carry a $U(1)'$ charge in this scenario, we will still refer to the Goldstone boson $J = \sqrt{2}\, \mathrm{Im}(\sigma) $ as a majoron.
	
	At one-loop level, one actually does obtain neutrino masses even for $\# \nu_L = \# N_R = \# S_R$; these are proportional to $\mu$, making it an inverse seesaw~\cite{Dev:2012sg}. We will omit a discussion of this interesting and rather minimal case, as it requires a calculation of the majoron couplings at loop level for consistency.
	
\end{enumerate}

All three cases have in common a heavy-neutrino mass submatrix of the form $\matrixx{\mu_1 & M\\ M^T & \mu_2}$ that does not commute with the coupling matrix of $\sigma$. In other words, the mass matrix consists of bare terms plus $U(1)'$-breaking terms, which ensures that $J$ will have ``flavor changing'' couplings, i.e.~off-diagonal couplings to the heavy states. This is the main difference compared to the singlet-majoron model, where these off-diagonal terms only arise at higher order in the seesaw expansion and are therefore very suppressed.
A further common feature of all three scenarios above is the existence of a pseudo-Dirac limit: for $\mu_j\ll M$, the heavy states will form quasi-degenerate pairs, which is precisely the situation of interest for our phase-space suppressed decays.
Below we will discuss the models in more detail.
The diagonalization of the mass matrices and calculation of the majoron couplings is described in detail in App.~\ref{sec:perturbation_theory}, here we will simply quote the results.

\subsection{Inverse Seesaw}
\label{sec:IS}

For simplicity we will for now work in the one-generational limit, i.e.~$\# \nu_L = \# N_R = \# S_R = 1$.
We assign $U(1)'$ charges $X(\nu_L) = X(N_R) =-X(\sigma)= 1$ and $X(S_R)=0$, allowing us to write the couplings of interest as
	\begin{align}
	\label{eq:iss_fundamental_lagrangian}
	\L \supset -\frac12 \overline{N}_L^c \matrixx{0 & m_D & 0 \\ m_D & 0 & \lambda \sigma^* \\ 0 & \lambda \sigma^* & \mu} N_L +\hc ,
	\end{align}
with $N_L = (\nu_L, N_R^c, S_R^c)$. Here we already replaced the SM scalar doublet by its vacuum expectation value to obtain the Dirac mass $m_D$, as in the standard seesaw mechanism. Upon $U(1)'$ symmetry breaking $\sigma \to \langle \sigma\rangle \equiv f/\sqrt{2}$, we obtain the IS mass matrix of Eq.~\eqref{eq:ISmassmatrix} with $M = \lambda f/\sqrt{2}$. For $m_D,\mu \ll M$, this gives three Majorana states $N_{1,2,3}$ with masses
\begin{align}
m_1 \simeq \mu\frac{m_D^2}{M^2}\,, &&
m_2 \simeq M - \frac{\mu}{2}\,, &&
m_3 \simeq M + \frac{\mu}{2}\,.
\end{align}
Taking $\mu$ to be positive without loss of generality, $N_3$ is the heaviest state, but almost degenerate with $N_2$. $N_1$ corresponds to the active-neutrino eigenstate with well-known IS mass proportional to $\mu$~\cite{Mohapatra:1986aw,Mohapatra:1986bd,GonzalezGarcia:1988rw,Frere:1989xb}.

The couplings of the  majoron $J$ to the neutrino mass eigenstates $N_j$ to lowest non-vanishing order are given by
	\begin{align}
	\L \ \supset \ -\frac{J}{2 f}\, \overline{N}_i \matrixx{-2\i \gamma_5 m_1 & \frac{m_D}{\sqrt{2}} & \i \gamma_5 \frac{m_D}{\sqrt{2}} \\   \frac{m_D}{\sqrt{2}} & \i\gamma_5 M & \frac{\mu}{2} \\ \i\gamma_5 \frac{m_D}{\sqrt{2}} & \frac{\mu}{2} & \i\gamma_5 M}_{ij} N_j \, .
	\label{eq:ISmajoron}
	\end{align}
Note that the off-diagonal coupling $J N_2 N_3$ of the quasi-degenerate states is proportional to the mass splitting $\mu \simeq m_3-m_2$.
The decay rates of interest to us can be immediately obtained with the formulae of App.~\ref{sec:majorana_decays}. In the limit $m_J \ll \mu \ll M$ they take the simple form
\begin{align}
\Gamma (N_3\to N_2 J) \simeq \frac{\mu^3}{8\pi f^2}\,, &&
\Gamma (N_{2,3}\to N_1 J) \simeq \frac{m_D^2 M}{32\pi f^2}\,, &&
\Gamma (J\to N_1 N_1) \simeq \frac{m_1^2 m_J}{4\pi f^2}\,.
\label{eq:ISdecays}
\end{align}
The last one corresponds to DM decay into active neutrinos and is of the same form as in the singlet-majoron model, i.e.~proportional to the neutrino-mass squared, but bigger by a factor of 4 due to the larger coupling. Since we assume sub-MeV DM masses, it will be incredibly difficult to directly search for such monochromatic neutrinos~\cite{Garcia-Cely:2017oco}, but one can still obtain a lower bound of $160$--$\unit[170]{Gyr}$ on this cold-DM decay from its cosmological impact~\cite{Audren:2014bca,Poulin:2016nat}. For a $\unit[7]{keV}$ majoron, this corresponds to a lower limit on the $U(1)'$ breaking scale $f>\unit[10^8]{GeV}$, assuming normal neutrino hierarchy.

It is convenient to split the discussion of DM production according to whether the production temperature $T_\text{prod} \sim M/5-M/3$ is above or below the electroweak phase transition of $\sim\unit[160]{GeV}$: above, we can set the electroweak vacuum expectation value to zero, which also turns off the decays $N_{2,3}\to N_1 J$; below, the neutrino masses are so small compared to $f>\unit[10^8]{GeV}$ that the scattering processes $N_{2,3} N_{2,3}\to JJ$ become negligible. More details are given below.

\subsubsection{DM production below the electroweak scale}

Dark matter is produced entirely below the EW phase transition for $M \lesssim 500$ GeV.
In this case, the optimal channel for cold DM production is $N_3\to N_2 J$, which gives a small average DM momentum $\langle x \rangle = \tfrac52 (1-m_2^2/m_3^2) \simeq 5 \mu/M \ll 1$. In order to make the competing $\langle x \rangle =5/2$ production channel $N_{2,3}\to N_1 J$, subdominant, we would naively need to impose the hierarchy $m_D^2 M \ll \mu^3$. This is not the usual hierarchy in IS, but poses no problems. 
We can be actually be more precise about this inequality; we have shown in Eq.~\eqref{eq:final_Ly_alpha_limit} that the quantity of interest for Lyman-$\alpha$ is $\sum_j\text{BR}_j \Delta_j^\eta$, so we should actually demand
\begin{align}
\frac{\Omega (N_{3}\to N_1 J)}{\Omega (N_3\to N_2 J)} \simeq \frac{\Gamma (N_{3}\to N_1 J)}{\Gamma (N_3\to N_2 J)} \simeq \frac{m_D^2 M}{4 \mu^3}
\stackrel{!}{<} \frac12\frac{\Delta_{N_3\to N_2 J}^\eta}{\Delta_{N_{3}\to N_1 J}^\eta}  \simeq \frac{2\mu^2}{M^2}
\label{eq:hot_decay_BR}
\end{align}
in order to forbid the $N_{2,3}\to N_1 J$ channels to contribute to the transfer function. In the last relation we have approximated $\eta \simeq 2$ for simplicity.
Eq.~\eqref{eq:hot_decay_BR} is a much stronger requirement than the naive guess $\Omega (N_{3}\to N_1 J)\stackrel{!}{<} \Omega (N_{3}\to N_2 J)$ and shows how dangerous even a small subcomponent of warm DM can be when it comes to structure formation. But if this relation $m_D^2 < 8\mu^5/M^3$ is satisfied, Eq.~\eqref{eq:final_Ly_alpha_limit} reduces back to the single decay case and the Lyman-$\alpha$ limit depends only on the $N_3\to N_2 J$ decay, i.e.~on $\langle x \rangle \simeq 5 \mu/M$, which can be made small.

Overall, our one-generation IS scenario has five parameters of interest, $\{m_J, \mu, m_D, M, f\}$, which have to reproduce the DM abundance and neutrino mass scale, as well as satisfy numerous inequalities to ensure compliance with Lyman-$\alpha$ and DM stability.
Fixing the DM abundance and neutrino mass $m_1$, we can express $\mu$ and $m_D$ in terms of $\{m_J, M, f\}$,
\begin{align}
\mu &\simeq  \unit[11]{GeV}\left(\frac{M}{\unit[300]{GeV}}\right)^{2/3}\left(\frac{f}{\unit[10^8]{GeV}}\right)^{2/3} \left(\frac{m_J}{\unit{keV}}\right)^{-1/3}  ,\label{eq:IS_mu}\\
m_D &\simeq  \unit[0.9]{MeV}\left(\frac{M}{\unit[300]{GeV}}\right)^{2/3}\left(\frac{f}{\unit[10^8]{GeV}}\right)^{-1/3} \left(\frac{m_J}{\unit{keV}}\right)^{1/6} \left(\frac{m_1}{\unit[0.1]{eV}}\right)^{1/2} ,
\end{align}
as well as determine the quantities
\begin{align}
\langle x\rangle &\simeq  0.18 \left(\frac{M}{\unit[300]{GeV}}\right)^{-1/3}\left(\frac{f}{\unit[10^8]{GeV}}\right)^{2/3} \left(\frac{m_J}{\unit{keV}}\right)^{-1/3} ,\\
\Gamma (J\to N_1 N_1) &\simeq  \frac{0.65}{\unit[170]{Gyr}}\left(\frac{f}{\unit[10^8]{GeV}}\right)^{-2} \left(\frac{m_J}{\unit{keV}}\right) \left(\frac{m_1}{\unit[0.1]{eV}}\right)^{2} ,\\
\frac{\Omega (N_{2,3}\to N_1 J)}{\Omega (N_3\to N_2 J)} &\simeq  9\times 10^{-8}\left(\frac{M}{\unit[300]{GeV}}\right)^{1/3}\left(\frac{f}{\unit[10^8]{GeV}}\right)^{-8/3} \left(\frac{m_J}{\unit{keV}}\right)^{4/3} \left(\frac{m_1}{\unit[0.1]{eV}}\right) .
\end{align}
The last quantity has to be smaller than $\tfrac{2}{25} \langle x \rangle^2$ (see Eq.~\eqref{eq:hot_decay_BR}) in order for $\langle x \rangle$ to be a reliable mean DM momentum that can be plugged into Eq.~\eqref{eq:translation2} to check consistency with Lyman-$\alpha$ constraints. 
We see that one can easily realize cold keV DM for heavy neutrinos in the $\unit[\mathcal{O}(100)]{GeV}$ range. Note that these heavy neutrinos decay fast, before Big-Bang nucleosynthesis, via the Yukawa coupling $\sim m_D/(246$ GeV). 

The decay $J \to \gamma \gamma$ via 2-loop neutrino-induced $J \to Z^* \to \gamma \gamma$ might be too slow to lead to an observable signature for  $m_D \sim $ MeV. However, other possibilities exist~\cite{Heeck:2017xbu}. For instance, a mixing with the Higgs boson $\sim 10^{-13}$ can easily lead to an observable flux, in particular for the putative 3.5 keV line. In any case, this would not affect the cold-enough production of dark matter. We stress here that this is true for all scenarios discussed below.

\subsubsection{DM production above the electroweak scale}

If the DM production temperature is above the electroweak scale, i.e.~$m_{2,3} >\unit{TeV}$, the decays $N_{2,3}\to N_1 J$ are forbidden by $SU(2)$ invariance. Nevertheless, there exist scattering processes that compete with our cold-DM production channel $N_3\to N_2 J$ that need to be taken into account to ensure that DM is cold enough.

If $M$ is close to $f$, the dominant DM production channel that competes with $N_3\to N_2 J$ comes from the scattering $N_j N_j \to J J$, seeing as the majoron couples most strongly diagonally to $N_{2,3}$. The cross section for this process in the limit of massless $J$ takes on the form
\begin{align}
\sigma (N_j N_j\to J J) = \frac{1}{128\pi} \frac{M^2}{f^4} \left(1-\frac{1}{\beta(s)^2}\right)\left(2\beta(s) + \log \left[\frac{1-\beta(s)}{1+\beta(s)}\right]\right) , \quad j = 2,3,
\label{eq:scattering}
\end{align}
with $\beta(s) = \sqrt{1-4 M^2/s}$. This matches the expression derived in Ref.~\cite{Gu:2009hn}; as shown there, this scattering process would thermalize $J$ if $f \lesssim 0.1 (M^3 M_\mathrm{Pl})^{\frac14}$. Even if this annihilation rate is too slow to reach equilibrium, it can still freeze-in a population of $J$~\cite{Frigerio:2011in} with mean momentum $\langle x \rangle = \mathcal{O}(3)$. To push DM below the keV scale, we therefore have to demand that this process gives only a subcomponent of DM. The DM abundance $\Omega$ can be readily calculated with the formulae from Ref.~\cite{Heeck:2017xbu}; numerically, the ratio can be approximated by
\begin{align}
\frac{\Omega (N_{2,3} N_{2,3}\to JJ)}{\Omega (N_3\to N_2 J)} \simeq 8\times 10^{-4} \frac{M^5}{f^4 \Gamma (N_3\to N_2 J)} \simeq 2\times 10^{-2} \frac{M^5}{f^2 \mu^3} \,,
\end{align}
which we demand to be smaller than $2\mu^2/M^2$ in order to not mess up the small mean DM momentum, by extrapolating the argument of the decay case~\eqref{eq:hot_decay_BR}.

For $M\ll f$, the scattering channels $N_{2,3} H \to L J$, $\bar L H \to \bar N_{2,3} J$, and $N_{2,3}\bar L \to  \bar H J$ can become relevant, as they are only suppressed by $(M/f)^2$. Similar to the scattering case above, we can calculate the resulting abundance $\Omega (N_{2,3} H L J)$ of rather warm DM from all these processes as
\begin{align}
\frac{\Omega (N_{2,3} H L J)}{\Omega (N_3\to N_2 J)} \simeq \frac{11 M^3 y_D^2}{192 \pi^3 f^2 \Gamma (N_3\to N_2 J)}  \,,
\end{align}
$y_D \equiv m_D/(\unit[246]{GeV})$ being the Yukawa coupling that leads to the neutrino Dirac mass $m_D$. We can once again express all relevant quantities in terms of $\{m_J, M, f\}$:
\begin{align}
\mu &\simeq  \unit[72]{GeV}\left(\frac{M}{\unit[5]{TeV}}\right)^{2/3}\left(\frac{f}{\unit[10^8]{GeV}}\right)^{2/3} \left(\frac{m_J}{\unit{keV}}\right)^{-1/3}  ,\\
m_D &\simeq  \unit[6]{MeV}\left(\frac{M}{\unit[5]{TeV}}\right)^{2/3}\left(\frac{f}{\unit[10^8]{GeV}}\right)^{-1/3} \left(\frac{m_J}{\unit{keV}}\right)^{1/6} \left(\frac{m_1}{\unit[0.1]{eV}}\right)^{1/2} ,\\
\langle x\rangle &\simeq  0.07 \left(\frac{M}{\unit[5]{TeV}}\right)^{-1/3}\left(\frac{f}{\unit[10^8]{GeV}}\right)^{2/3} \left(\frac{m_J}{\unit{keV}}\right)^{-1/3} ,\\
\Gamma (J\to N_1 N_1) &\simeq  \frac{0.65}{\unit[170]{Gyr}}\left(\frac{f}{\unit[10^8]{GeV}}\right)^{-2} \left(\frac{m_J}{\unit{keV}}\right) \left(\frac{m_1}{\unit[0.1]{eV}}\right)^{2} ,\\
\frac{\Omega (N_{2,3} N_{2,3}\to JJ)}{\Omega (N_3\to N_2 J)} &\simeq  1.6\times 10^{-5}\left(\frac{M}{\unit[5]{TeV}}\right)^{3}\left(\frac{f}{\unit[10^8]{GeV}}\right)^{-4} \left(\frac{m_J}{\unit{keV}}\right) ,\\
\frac{\Omega (N_{2,3} H L J)}{\Omega (N_3\to N_2 J)} &\simeq 9\times 10^{-6}\left(\frac{M}{\unit[5]{TeV}}\right)^{7/3}\left(\frac{f}{\unit[10^8]{GeV}}\right)^{-8/3} \left(\frac{m_J}{\unit{keV}}\right)^{4/3}  \left(\frac{m_1}{\unit[0.1]{eV}}\right) .
\end{align}
In addition, we should demand $m_D \lesssim \unit[250]{GeV}$ and $M \lesssim f$ in order to avoid having non-perturbatively large Yukawa couplings in the model, but this turns out to be a much weaker constraint than the above inequalities. Notice, however, that these Yukawa couplings are anyway large enough to guarantee the thermalization of $N_{2,3}$ with the SM bath, as assumed in the analysis of Sec.~\ref{sec:freeze-in}.

For a given DM mass, $\Gamma (J\to N_1 N_1)$ provides a lower bound on $f$ to ensure DM stability. $\langle x\rangle$ together with Lyman-$\alpha$ then gives a lower bound on $M$, which at some point leads to a large DM production rate via $N_{2,3} N_{2,3}\to JJ$. From the above it is clear that $\mathcal{O}(\unit{TeV})$ right-handed neutrinos can generate cold keV DM without running into problems with Lyman-$\alpha$. 

\subsubsection*{}

The above shows that the inverse seesaw mechanism can not only generate light neutrino masses, but also provide the necessary ingredients to produce $\mathcal{O}(\unit{keV})$ cold majoron DM. Aside from the unavoidable decay channel of DM into neutrinos, additional interactions could lead to a detectable decay into $\gamma\gamma$ without endangering our production mechanism~\cite{Heeck:2017xbu}.
So far we have only considered the one-generational IS, but the above discussion can be generalized straightforwardly.
Let us remark on one particularly interesting consequence of more generations, that is lepton flavor violation. The IS with hierarchy $\mu\ll m_D\ll M$ has the feature of keeping neutrino masses small without tiny active--sterile mixing angles. This makes it a popular model to discuss rare flavor violating decays such as $\mu\to e\gamma$, induced at one-loop level by the not-too-heavy neutrinos. This feature is diluted when adopting our hierarchy of interest, $m_D\ll \mu\ll M$, which looks more like a normal seesaw mechanism when it comes to active--sterile mixing. Due to possible matrix cancellations in the three-generational case it is of course still possible to have detectable rates for e.g.~$\mu\to e\gamma$, but there is little predictivity from the DM side. Rare decays involving majorons, e.g.~$\mu\to e J$, are similarly expected to be suppressed if $J$ forms DM, but this might be discussed elsewhere in more detail. Finally, notice that whereas in principle the quasi-degenerate states $N_{2,3}$ could be used to generate the baryon asymmetry of the Universe  via resonant leptogenesis~\cite{Pilaftsis:1997jf,Pilaftsis:2003gt,Dev:2014laa} (if their mass is above the weak scale), or via the Akhmedov--Rubakov--Smirnov~\cite{Akhmedov:1998qx,Asaka:2005pn}  and Higgs-decay mechanisms~\cite{Hambye:2016sby} (if below), their mass splitting, essentially given by $\mu$, turns out to be too large (see~\eqref{eq:IS_mu}) to have a successful generation of the asymmetry.

\subsection{Extended Inverse Seesaw}
\label{sec:EIS}

The second model of interest is the Extended Inverse Seesaw, which we obtain by assigning $X(\nu_L) = X(N_R) = 1$, $X(\sigma)=2$, and $X(S_R)=-1$, leading to the allowed couplings for one generation $N_L = (\nu_L, N_R^c, S_R^c)$
	\begin{align}
	\L \supset -\frac12 \overline{N}_L^c \matrixx{0 & m_D & 0 \\ m_D & \lambda_1 \sigma & M \\ 0 & M & \lambda_2 \sigma^*} N_L +\hc ,
	\end{align}
which leads to the EIS mass matrix of Eq.~\eqref{eq:EISmassmatrix} upon $U(1)'$ symmetry breaking, where $\mu_j = \lambda_j \langle \sigma\rangle = \lambda_j f/\sqrt{2}$.
Taking all mass terms to be real and positive for simplicity,\footnote{A more general expression allowing for complex $\mu_1$ is given in Eq.~\eqref{eq:EISmajoron_general}.} with hierarchy $\mu_j, m_D\ll M$, the mass spectrum is
\begin{align}
m_1\simeq \mu_2 \frac{m_D^2}{M^2}\,, &&
m_2\simeq M - \frac{\mu_1}{2}- \frac{\mu_2}{2}\,,&&
m_3\simeq M + \frac{\mu_1}{2}+ \frac{\mu_2}{2}\,,
\end{align}
and the majoron couplings to lowest order are
\begin{align}
\L \supset -\frac{J}{2f}\overline{N_i}\begin{pmatrix}
        \i\gamma_5 m_1 & -\i\gamma_5\frac{\mu_2 m_D}{\sqrt{2}M} & -\frac{\mu_2 m_D}{\sqrt{2}M}\\
        -\i\gamma_5\frac{\mu_2 m_D}{\sqrt{2}M} & -\i\gamma_5\frac{\mu_1 - \mu_2}{2} & \frac{\mu_1 + \mu_2}{2}\\
         -\frac{\mu_2 m_D}{\sqrt{2}M} & \frac{\mu_1 + \mu_2}{2} & \i\gamma_5\frac{\mu_1 - \mu_2}{2}\\
    \end{pmatrix}_{ij}N_j.
\label{eq:EISmajoron}
\end{align}
Compared to the IS case from above, none of the couplings here are large, but rather of order $\mu_j/f$ or even further suppressed by powers of $m_D/M$. As a result, the decay $N_3\to N_2 J$ automatically dominates the DM production compared to $N_{2,3}\to N_1 J$ (suppressed by $m_D/M$) and $N_{2,3}N_{2,3}\to J J$ (suppressed by $2\to 2$ phase space). 
The decays relevant for DM production and decay are
\begin{align}
\Gamma (N_3\to N_2 J) \simeq \frac{(\sum_j \mu_j)^3}{8\pi f^2}\,, &&
\Gamma (N_{2,3}\to N_1 J) \simeq \frac{\mu_2^2 m_D^2}{32\pi M f^2}\,, &&
\Gamma (J\to N_1 N_1) \simeq \frac{m_1^2 m_J}{16\pi f^2}\,.
\label{eq:EISdecays}
\end{align}
Fixing the relic abundance with Eq.~\eqref{eq:massless_approx}, we can express $\{m_J,\mu_1,\mu_2,m_D,M,f\}$ in terms of $\{m_1, \tau (J\to N_1 N_1),\mu_1,\mu_2\}$ as
\begin{align}
m_D &\simeq \unit[0.2]{GeV}\left(\frac{\unit[170]{Gyr}}{\tau_J}\right)^{1/2}\left(\frac{m_1}{\unit[0.1]{eV}}\right)^{-1/2}\left(\frac{\sum_j \mu_j}{\unit{TeV}}\right)^{3/2}\left(\frac{\mu_2}{\unit{TeV}}\right)^{-1/2},\\
M &\simeq \unit[6\times 10^5]{GeV}\left(\frac{\unit[170]{Gyr}}{\tau_J}\right)^{1/2}\left(\frac{m_1}{\unit[0.1]{eV}}\right)^{-1}\left(\frac{\sum_j \mu_j}{\unit{TeV}}\right)^{3/2},\label{eq:EIS_M}\\
\langle x \rangle &\simeq 8\times 10^{-3}\left(\frac{\unit[170]{Gyr}}{\tau_J}\right)^{-1/2}\left(\frac{m_1}{\unit[0.1]{eV}}\right)\left(\frac{\sum_j \mu_j}{\unit{TeV}}\right)^{-1/2}.
\end{align}
For a fixed neutrino mass $m_1= \unit[0.1]{eV}$ we can push $\mu_j$ as high as $\unit[10^6]{GeV}$ before $m_D$ reaches values that would imply non-perturbative Yukawa couplings, which allows us to go down to $\langle x \rangle \sim 2\times 10^{-4}$. We can thus easily have DM masses around keV without violating the Lyman-$\alpha$ bound from Eq.~\eqref{eq:final_Ly_alpha_limit}.
Finally, notice that in this case leptogenesis via the decay of $N_{2,3}$ appears to be possible: for the limit values $\mu_j \sim \unit[10^3]{TeV}$, eq.~\eqref{eq:EIS_M} gives $m_{N_{2,3}} \sim \unit[10^{10}]{GeV}$, in which case leptogenesis is known to be possible even without  mass degeneracy (which is however present).

\subsection{Extended Seesaw}
\label{sec:ES}

The last model with a promising coupling structure for $A\to B\,\text{DM}$ decays is the Extended Seesaw, obtained by assigning $X(\nu_L) = X(N_R) = 0$, and $X(\sigma)= X(S_R)=1$. Taking again a simple one-generation model $\# \nu_L = \# S_R = 1$, we need $\# N_R = 2$ to generate enough mass terms at tree level. The minimal model is then
	\begin{align}
	\L \supset -\frac12 \overline{N}_L^c \matrixx{0 & m_{D,1} & m_{D,2} & 0 \\ m_{D,1} & \mu_{11} & \mu_{12} & 0 \\ m_{D,2} & \mu_{12} & \mu_{22} & \lambda \sigma \\ 0 & 0 & \lambda \sigma & 0} N_L +\hc ,
	\end{align}
	with $N_L = (\nu_L, N_{R,1}^c, N_{R,2}^c, S_R^c)$. We have already performed a rotation of $N_{R,1}$ and $N_{R,2}$ to eliminate a second possible $S_R$ coupling. Upon symmetry breaking, $M = \lambda \langle \sigma \rangle = \lambda f/\sqrt{2}$, we obtain four massive Majorana fermions by diagonalizing the above mass matrix.
	
	Let us first consider the case with $m_{D,2} = \mu_{12} = 0$, which implies that $(\nu_L, N_{R,1}^c)$ and $(N_{R,2}^c, S_R^c)$ decouple and do not mix. The masses are then simply
	\begin{align}
	m_1 \simeq \frac{m_{D,1}^2}{\mu_{11}}\,, &&
	m_2 \simeq \mu_{11}\,, &&
	m_3 \simeq M - \frac{\mu_{22}}{2}\,, &&
	m_4 \simeq M + \frac{\mu_{22}}{2}\,, &&
	\end{align}
	assuming all entries real, positive, and with hierarchies $m_{D,1} \ll \mu_{11}$ and $\mu_{22}\ll M$. The majoron $J$ only couples to $N_3$ and $N_4$ in this approximation, with the DM-production decay rate
	\begin{align}
	\Gamma (N_4\to N_3 J) \simeq \frac{\mu_{22}^3}{8\pi f^2} \,,
	\end{align}
	as well as scattering rate $N_{3,4}N_{3,4}\to J J$ given by Eq.~\eqref{eq:scattering}.
	Similar to the IS case, we can easily go down to $m_J\sim \unit{keV}$ without violating Lyman-$\alpha$ constraints before the scattering production starts to dominate.
	To make $N_3$ unstable, we need a small deviation from $m_{D,2} = \mu_{12} = 0$ in order to induce mixing among the neutral fermions, but this will not change our discussion much. These couplings, as well as the ones present in the scalar sector, could also be responsible for the thermalization of $N_4$. Notice that this is the only model in which the active neutrino masses are completely decoupled from the DM properties, including the absence of the DM decay channel $J\to N_1 N_1$. This channel will however open up once we deviate from $m_{D,2} = \mu_{12} = 0$.
	We omit a discussion of the general case due to the large number of free parameters, but it should be clear from the case above that this is indeed a valid model to obtain keV DM.
	The simpler case with $\# \nu_L = \# S_R = \# N_R$, where neutrino masses only arise at loop level~\cite{Dev:2012sg}, seems promising but goes beyond the scope of this article.

\section{Conclusion}
\label{sec:conclusion}

Feebly coupled light dark matter poses an interesting alternative to the more commonly studied electroweak-scale inspired candidates. A lower bound on the DM mass can be inferred from the existence of small-scale structures in our Universe, for example by studies of the Lyman-$\alpha$ forest. Such limits are usually dependent on the full DM distribution function, and hence the production mechanism.
In this article we have studied DM freeze-in via two-body decays $A\to B\,\mathrm{DM}$, with heavy $A$ in equilibrium with the SM. For $m_\mathrm{DM}\ll m_A-m_B\ll m_A$, the decay is phase-space suppressed and leads to cold light DM. As verified with the linear DM power spectrum, this makes it possible to produce bosonic DM with masses around keV without violating Lyman-$\alpha$ constraints. 
To assess the viability of cold \emph{sub-}keV DM would require the full solution to the integro-differential Boltzmann equations beyond the freeze-in approximation.

The required ingredients for this scenario, light bosonic DM coupled off-diagonally to heavy quasi-degenerate particles, can naturally be found in extended seesaw mechanisms with spontaneously broken lepton symmetries. DM then arises as the pseudo-Goldstone of a $U(1)'$ symmetry and couples to quasi-degenerate right-handed neutrinos. As a bonus, these models automatically induce small masses for the active neutrinos, and might even lead to leptogenesis.
As illustrated within the one-generational approximation, DM masses down to keV can be realized in all cases considered without violating Lyman-$\alpha$ constraints. 

Finally, these models could lead to observable signatures in the x-ray spectrum coming from DM decay. In particular, the putative 3.5 keV line could be generated, at the required strength, by the pseudo-Goldstone decay $J \to \gamma \gamma$ without spoiling the cold-enough production of dark matter.

\section*{Acknowledgements}
We thank Laura Lopez Honorez for useful discussions and the referee for important comments.
SB and AN thank the theory group of ULB for hospitality during the course of this work.
JH is a postdoctoral researcher of the F.R.S.-FNRS; DT is supported by a ULB postdoctoral fellowship and the Belgian Federal Science Policy (IAP P7/37).

\appendix

\section{Decays with Majorana fermions}
\label{sec:majorana_decays}

For completeness we list relevant decay widths involving Majorana fermions. We start with a real scalar particle $\phi$ coupled to a Majorana fermion $N = N^c$ via
\begin{align}
\L = \tfrac12 \del_\mu\phi\del^\mu \phi  - \tfrac12 m_\phi^2 \phi^2 + \tfrac12 \overline{N}(\i \slashed{\del}-m_N )N - \phi \overline{N} (y_S + y_P \i \gamma_5)N
\end{align}
with real couplings $y_{S,P}$. For $2 m_N < m_\phi$, this gives the decay rate
\begin{align}
\Gamma (\phi\to NN) = \frac{1}{4\pi m_\phi^2} \left[y_S^2 (m_\phi^2-4m_N^2) + y_P^2 m_\phi^2\right] \left(m_\phi^2 - 4 m_N^2\right)^{1/2} .
\end{align}
Next, let us couple the scalar to two different Majorana fermions $N_{1,2}$:
\begin{align}
\L = \frac12 \del_\mu\phi\del^\mu \phi- \frac12 m_\phi^2 \phi^2 + \frac12 \sum_{j=1,2}\overline{N}_j(\i \slashed{\del}-  m_j)N_j  - \left(\phi \overline{N}_1 (y_S + y_P \i \gamma_5)N_2 +\hc\right) .
\end{align}
Notice that we have added the hermitian conjugate to the interaction term and allowed for complex $y_{S,P}$, which will introduce a factor of 2 in the amplitude, see for example Refs.~\cite{Haber:1984rc,Denner:1992vza}. The different decay rates, if kinematically accessible, are then
\begin{align}
\Gamma (\phi\to N_1 N_2) &= \frac{1}{2\pi m_\phi^3} \left[|y_S|^2 (m_\phi^2-(m_1+m_2)^2) + |y_P|^2 (m_\phi^2-(m_1-m_2)^2)\right] \notag\\
 &\quad\times\left[ (m_\phi^2 -(m_1+m_2)^2)(m_\phi^2 -(m_1-m_2)^2)\right]^{1/2} \\
\Gamma (N_1 \to N_2 \phi) &= \frac{1}{4\pi m_1^3} \left[|y_S|^2 ((m_1+m_2)^2-m_\phi^2) + |y_P|^2 ((m_1-m_2)^2-m_\phi^2)\right] \notag\\
 &\quad\times\left[ (m_1^2 -(m_\phi+m_2)^2)(m_1^2 -(m_\phi-m_2)^2)\right]^{1/2} .
\end{align}
$\Gamma (N_2 \to N_1 \phi)$ simply follows from $\Gamma (N_1 \to N_2 \phi)$ by a change of indices.

\section{Derivation of Boltzmann equation}
\label{sec:boltzmann_derivation}

In this appendix, we give further details on the derivation of the Boltzmann equation for DM given in Eq.~\eqref{eq:fullMBdistribution}.
Assuming an early production of DM and neglecting the change of $g_*$ during DM production time, the equation governing the phase-space distribution ${f_{\mathrm{DM}}(p) \equiv f_{\mathrm{DM}}(\mathbf{|p|})} $ for DM is
\begin{equation}
      \dfrac{\partial f_{\mathrm{DM}}}{\partial t} - H p \dfrac{\partial f_{\mathrm{DM}}}{\partial p} = \mathcal{C} (p)\,,
      \label{eq:general_boltzmann}
\end{equation}
$H(T) = T^2/M_0$ being the Hubble constant in the radiation-dominated epoch with $M_\text{Pl}$ being the Planck mass and ${M_0 \simeq M_\text{Pl}/(1.66 \, g_*^{1/2})}$.
Neglecting quantum effects, the collision term for the process $A \to B \, \mathrm{DM}$ reads:
\begin{equation}
 \mathcal{C}(p) = \dfrac{1}{2 E} \int \dfrac{\mathrm{d}^3 \mathbf{p}_A}{(2 \pi)^3 \, 2 E_{A}} \int   \dfrac{\mathrm{d}^3 \mathbf{p}_B}{(2 \pi)^3 \, 2 E_{B}} (2 \pi)^4 \, \delta^{(4)}\left( p_A - p_B - p \right) |\mathcal{M}|^2 \; f_A(E_A(p_A))\,,
 \label{eq:collisionTerm}
\end{equation}
where $|\mathcal{M}|^2$ is the Lorentz invariant squared matrix element summed over both initial and final degrees of freedom. This expression still holds in the case where $B = \mathrm{DM}$. 

The cases of both massless daughter particles and massless DM/massive secondary particle have been studied in Ref.~\cite{Petraki:2007gq} and Ref.~\cite{Heeck:2017xbu}, respectively. Although highly motivated in the case where the mother particle's mass is of order $\unit[100]{GeV}$ and expected DM mass at $7$ keV, the massless DM approximation may potentially yield incorrect results in some regions of the parameter space, so we will keep $m_\text{DM}\neq 0$.

Computing Eq.~\eqref{eq:collisionTerm} for massive final state particles, one can get rid of the spatial Dirac-$\delta$ by integrating over $|\mathbf{p_B}|$ under isotropy assumption. Choosing the reference axis defining angles along $\mathbf{p}$ and calling $\theta = \sphericalangle (\mathbf{p}, \mathbf{p}_A)$, the remaining integral reads:
\begin{equation}
\begin{split}
    \mathcal{C}(p) = & \dfrac{1}{16 \pi} \dfrac{1}{E}  \int_{p_A} \int_{\cos \theta} \dfrac{\mathrm{d} p_A \, p_A^2 \: \mathrm{d}\cos \theta}{\sqrt{p_A^2 + m_A^2}} \dfrac{1}{\sqrt{m_{B}^2 + p^2 + p_A^2 - 2\, p\, p_A \cos \theta}}   \\
    & \times \delta \left( E + \sqrt{m_{B}^2 + p^2 + p_A^2 - 2 \,p \, p_A \cos \theta} - \sqrt{p_A^2 + m_A^2} \right) 
     |\mathcal{M}|^2 \; f_A(p_A)\,.
\end{split}
\end{equation}
Performing the integration over $\cos \theta$ and calling 
\begin{equation}
\cos \theta^* = \dfrac{m_B^2 + p^2 + p_A^2 - (\sqrt{m_A^2 + p_A^2} - E)^2}{2 \, p \, p_A},
\end{equation} 
the collision term becomes
\begin{equation}
\begin{split}
    \mathcal{C}(p) = \dfrac{|\mathcal{M}|^2}{16 \pi} \dfrac{1}{E \, p}  \int_{p_A}  \dfrac{\mathrm{d} p_A \, p_A \,  f_A(p_A)}{\sqrt{p_A^2 + m_A^2}}  \,,
\end{split}
\end{equation}
provided that 
\begin{equation}
    |\cos \theta^*| \leq 1
    \label{eq:kinmtCond}
\end{equation}
for the collision term not to vanish. This condition affects the boundaries of the integral over $p_A$. Using energy conservation and the normalization of the energy--momentum four-vector, Eq.~\eqref{eq:kinmtCond} yields:
\begin{equation}
     E_A^2 - E_A \dfrac{\Lambda \, E}{m_{\mathrm{DM}}^2} + \dfrac{\Lambda^2 + 4\, p^2 \, m_A^2}{4 \, m_{\mathrm{DM}}^2} \leq 0
\end{equation}
with $\Lambda \equiv m_A^2 + m_{\mathrm{DM}}^2 - m_B^2 \geqslant 0$. This can only be possible if
\begin{equation}
     (m_A^2-(m_B+m_\mathrm{DM})^2)(m_A^2-(m_B-m_\mathrm{DM})^2)  \geqslant 0\,.
		\label{eq:kinematic_inequality}
\end{equation}
Eq.~\eqref{eq:kinematic_inequality} is always fulfilled if the decay is kinematically allowed, so we obtain the boundaries
\begin{equation}
    E_A^{\pm} = \dfrac{1}{2} \left[ \dfrac{(m_A^2 + m_{\mathrm{DM}}^2 - m_B^2)}{m_{\mathrm{DM}}^2} E  \pm \dfrac{p \sqrt{(m_A^2-(m_B+m_\mathrm{DM})^2)(m_A^2-(m_B-m_\mathrm{DM})^2)} }{m_{\mathrm{DM}}^2} \right].
\end{equation}
The collision term then eventually reads:
\begin{equation}
\mathcal{C}(p) = \dfrac{|\mathcal{M}|^2}{\, 16 \pi \,E \, p \,} \displaystyle \int_{E_{A}^-}^{E_{A}^+} \mathrm{d} E_A \, f_A(E_A)\,.
\end{equation}

Switching variables $x \equiv p/T_{\text{bath}}$, $r \equiv m_A/T_{\text{bath}}$, $\xi = E_A/T_{\text{bath}}$ with $f_{\mathrm{DM}}(p(T)) = f_{\mathrm{DM}}(x,r)$ and recalling our working assumption that $g_{*}$ stays constant during the DM production epoch, Eq.~\eqref{eq:general_boltzmann} simplifies into
\begin{equation}
\dfrac{m_A^2}{M_0 \, r} \dfrac{\partial f_\mathrm{DM}}{\partial r} = \mathcal{C}(x,r)\,,
\end{equation}
where we use once again the radiation-dominated epoch at the time of production of DM. In the end, using the notation introduced in Eq.~\eqref{eq:pDM} and expressing $|\mathcal{M}|^2$ in terms of the partial decay width $\Gamma$ according to the convention used in Ref.~\cite{Heeck:2017xbu}, the Boltzmann equation governing DM distribution function simplifies into:
\begin{equation}
\dfrac{\partial f_\mathrm{DM}}{\partial r} 
= \dfrac{g_A \, S\,  \Gamma \, M_0 \, r^2}{2 \, p_{\mathrm{DM}} \, x \, \sqrt{m_A^2 x^2 + m_{\mathrm{DM}}^2 r^2}} \displaystyle \int_{\xi^-}^{\xi^+} \mathrm{d} \xi \, f_A(\xi)\,,
\end{equation}
where $S$ is the symmetry factor, $p_{\mathrm{DM}}$ the DM momentum in the rest-frame of $A$ (Eq.~\eqref{eq:pDM}) and $\xi^\pm = E_A^\pm/T$.
Assuming a Maxwell--Boltzmann distribution function for $A$, $f_A(\xi) = e^{-\xi}$, directly leads to the kinetic equation Eq.~\eqref{eq:fullMBdistribution}.

\section{Perturbation theory for the extended seesaw}
\label{sec:perturbation_theory}

When dealing with majoron DM produced with the mechanisms described above, one invariably has to write a Takagi factorization for the seesaw mass matrix; this is necessary to obtain both the neutrino mass eigenstates and their couplings to the majoron. For instance, one can show that the fundamental Lagrangian for the inverse seesaw of Eq.~\eqref{eq:iss_fundamental_lagrangian} essentially splits into Majorana mass terms
\begin{align}
    \L \supset -\frac{1}{2}m_i\overline{N}_i N_i 
\end{align}
and terms coupling the neutrino mass eigenstates $N_i$ to the majoron $J$,
\begin{align}
    \mathcal{L}_{\textrm{int}} \supset -\frac{\i J}{f}M\overline{N}_i N_j\left[\gamma_5\Re\left(\frac{U_{2i}U_{3j} + U_{2j}U_{3i}}{2}\right) + \i\Im\left(\frac{U_{2i}U_{3j} + U_{2j}U_{3i}}{2}\right)\right],
\end{align}
where $U$ is a unitary matrix such that
\begin{align}
    U^\dagger\begin{pmatrix}
        0 & m_D & 0\\
        m_D & 0 & M\\
        0 & M & \mu
    \end{pmatrix}\overline{U} = \diag(m_1, m_2, m_3),
\end{align}
and $m_i \in \mathbf{R}_+$. This operation on the inverse seesaw mass matrix is known as a \textit{Takagi factorization}. Thus, to find the masses $m_i$ and the coupling of the mass eigenstates to the majoron, one needs to estimate the relevant entries of $U$. Unfortunately, Takagi factorization is similar to diagonalization and in general, we do not know any closed form. Therefore, we develop in this appendix a theory allowing to estimate the entries of $U$ by a \textit{perturbative} method.

\subsection{Preliminaries}

\begin{definition}
    Let $M$ be a complex $n \times n$ symmetric matrix. A singular vector $x$ is a vector of $\mathbf{C}^n$ such that $M\overline{x} \in \mathbf{C}x$. If $M\overline{x} = \lambda x$, one says that $\lambda$ is the singular value associated to the singular vector $x$.
\end{definition}

\begin{remark}
    That $x$ is a singular vector of $M$ with singular value $\lambda$ implies that $x$ (respectively $\overline{x}$) is an eigenvector of $M\overline{M}$ (respectively $\overline{M}M$) with eigenvalue $|\lambda|^2$. If the eigenvalues of $M\overline{M}$ are non-degenerate, one can use this remark as well as the theorem to conclude that any eigenvector of $M\overline{M}$ is a singular vector of $M$.
\end{remark}

The theorem below introduces the Takagi factorization that will be used extensively.

\begin{theorem}{\emph{Takagi factorization:}}
    Let $M$ be a complex symmetric $n \times n$ matrix. Then there is a unitary matrix $U$ such that $U^\dagger M\overline{U} = \Delta$ where $\Delta$ is diagonal real positive.
\end{theorem}

\begin{remark}
    In the theorem above, the columns of $U$ are singular vectors of $M$. In physics, when the $M$ represents a mass matrix, we call the singular vectors mass eigenstates and the diagonal elements of $\Delta$ the masses.
\end{remark}

There is a closed form for the Takagi factorization of complex $2 \times 2$ matrices that works for most matrices. Let $M = \begin{pmatrix}
    a & c\\
    c & d
\end{pmatrix}$ be such a matrix and define
\begin{align}
    t & \equiv \frac{|a|^2 - |d|^2}{\sqrt{4|\overline{a}c + \overline{c}d|^2 + (|a|^2 - |d|^2)^2}}\,, & e^{\i\varphi}  & \equiv \frac{\overline{a}c + \overline{c}d}{|\overline{a}c + \overline{c}d|}\,,\\
    e^{\i\psi} & \equiv \frac{a\sqrt{1 + t}e^{\i\varphi} + c\sqrt{1 - t}}{\left|a\sqrt{1 + t}e^{\i\varphi} + c\sqrt{1 - t}\right|}\,, &
    e^{\i\psi'} & \equiv \frac{a\sqrt{1 - t}e^{\i\varphi} - c\sqrt{1 + t}}{\left|a\sqrt{1 - t}e^{\i\varphi} - c\sqrt{1 + t}\right|}\,,
\end{align}
where we assume all denominators to be non-zero. The unitary matrix $U$ 
\begin{align}
    \label{eq:takagi_22_closed_form}
    U & \equiv \frac{1}{\sqrt{2}}\begin{pmatrix}
        e^{\i\psi/2}\sqrt{1 + t}e^{-\i\varphi/2} & e^{\i\psi'/2}\sqrt{1 - t}e^{-\i\varphi/2}\\
        e^{\i\psi/2}\sqrt{1 - t}e^{\i\varphi/2} & -e^{\i\psi'/2}\sqrt{1 + t}e^{\i\varphi/2}
    \end{pmatrix}
\end{align}
then diagonalizes $U^\dagger M\overline{U} = \diag(\lambda_1, \lambda_2)$ with positive singular values
\begin{align}
    \lambda_j & = \sqrt{\frac{1}{2}\left(|a|^2 + |d|^2 + 2|c|^2\right) + (-1)^{j-1} \frac{1}{2}\sqrt{4|\overline{a}c + \overline{c}d|^2 + (|a|^2 - |d|^2)^2}}\,.
\end{align}
There is unfortunately no simple algebraic form for the Takagi factorization for larger matrices, leading us to perturbative Takagi factorization (or perturbation of singular vectors): we start from a symmetric matrix we can explicitly factorize, perturb it with another symmetric matrix and try to figure out how the singular vectors and singular values are affected. To understand the theory of perturbation for singular vectors, it is worth first considering the more general problem of singular subspace perturbation.

\subsection{General theory of perturbation for singular subspaces}
\label{sec:perturbation_singular_subspaces}

In the following, we develop a theory that allows to approximate with controlled error the singular vectors of a given symmetric matrix. The general framework in which these approximations are derived is that of  singular subspace perturbation. This is similar to invariant subspace perturbation, which is extensively discussed in Ref.~\cite{stewart_sun_2004}. In this paragraph, we adapt some of the very general results of~\cite{stewart_sun_2004} to deal with singular subspaces instead of invariant subspaces. In all the rest of the appendix, we apply these general results to derive approximations for singular vectors in various cases of interest.

We start with the definition of a singular subspace, which generalizes that of a singular vector:
\begin{definition}
    Let $M$ be a symmetric $n \times n$ complex matrix. A singular subspace $E$ is $\mathbf{C}$-subspace of $\mathbf{C}^n$ such that $M\overline{E} \subset E$.
\end{definition}
In general, the singular subspace perturbation problem consists to find a unitary matrix close to identity that block-factorizes:
\begin{align}
    \label{eq:general_perturbation_problem}
    M & = \begin{pmatrix}
        L_1 & G\\
        G^T & L_2
    \end{pmatrix},
\end{align}
where the blocks $L_1$ and $L_2$ are symmetric and $G$ is ``small''. That is, we want to find $U$ unitary such that
\begin{align}
    \label{eq:singular_subspace_perturbation_U}
    U^\dagger M\overline{U} & = \begin{pmatrix}
        L_1' & 0\\
        0 & L_2'
    \end{pmatrix},
\end{align}
where $L_1', L_2'$ are symmetric matrices that we expect to be close to $L_1, L_2$ since $G$ is small.
\begin{remark}
    The reason why it is called singular subspace perturbation is the following. Suppose $L_1$, $L_2$ have respective sizes $n_1 \times n_1$, $n_2 \times n_2$. Then in case $G = 0$, $\varepsilon_1 = (1, 0, \ldots), \varepsilon_2, \ldots, \varepsilon_{n_1}$ obviously span a singular subspace for $M$ and $\varepsilon_{n_1 + 1}, \ldots, \varepsilon_{n_1 + n_2}$ span another singular subspace. Now, if $G$ is small, Eq.~\eqref{eq:singular_subspace_perturbation_U} simply says that the first $n_1$ columns of $U$ (that we expect to be close to identity) span a singular subspace of $M$ and the last $n_2$ another singular subspace. Now, suppose we know a singular subspace $E$ of dimension $n_1$ for some matrix $M_0$ and we want to know ``how this singular subspace changes'' for $M_0 + \Delta M$ where $\Delta M$ is sufficiently small. If one defines a unitary matrix $U_0$ whose first $n_1$ columns span $E$, then
    \begin{align}
        U_0^\dagger M_0\overline{U_0} & = \begin{pmatrix}
            \widetilde{L_1} & 0\\
            0 & \widetilde{L_2}
        \end{pmatrix}
    \end{align}
    is block-diagonal, where block $\widetilde{L_1}$ has size $n_1 \times n_1$ and block $\widetilde{L_2}$ has size $n_2 \times n_2$. Since $\Delta M$ is small, we expect $U_0^\dagger (M_0 + \Delta M)\overline{U_0}$ to have only small off-diagonal blocks. Thus by solving the general problem stated above for $M = U_0^\dagger (M_0 + \Delta M)\overline{U_0}$, the perturbated $E$ singular subspace is simply given by the $n_1$ first columns of $U_0U$.
\end{remark}

The general procedure is to look for $U$ in the form of an ansatz
\begin{align}
    U & = \begin{pmatrix}
        (1 + \xi\xi^\dagger)^{-1/2} & \xi(1 + \xi^\dagger\xi)^{-1/2}\\
        -\xi^\dagger(1 + \xi\xi^\dagger)^{-1/2} & (1 + \xi^\dagger\xi)^{-1/2}
    \end{pmatrix},
\end{align}
where $\xi$ is a ``small'' matrix. What we want is simply to cancel the $(1, 2)$ block of $U^\dagger M\overline{U}$, which translates to the following equation for $\xi$:
\begin{align}
    \label{eq:general_xi_definition}
    G + L_1\overline{\xi} - \xi L_2 - \xi G^T\overline{\xi} = 0.
\end{align}
Denote by $T$ the operator $\xi \longmapsto \xi L_2 - L_1\overline{\xi}$ and assume it is invertible. One can then attempt an iterative resolution in the following way: set $\xi^{(0)} = 0$ and for $n \geq 0$, $\xi^{(n + 1)} = T^{-1}(G - \xi^{(n)}G^T\overline{\xi^{(n)}})$. This is guaranteed to converge for sufficiently small $G$. More precisely, the following general theorem allows to bound the error at every iteration:

\begin{theorem}
    \label{th:general_xi_equation_theorem}
    If  $\rho \equiv 4\lVert T^{-1}(G)\rVert\lVert T^{-1} \rVert\lVert G \rVert < 1$,\footnote{$\lVert T^{-1} \rVert$ refers to the operator norm of $T^{-1}$ with respect to the usual $2$-norm.} then the iterative procedure above converges. In addition, the following inequalities hold:
    \begin{align}
        \lVert \xi^{(n)} \rVert & \leq 2\lVert T^{-1}(G)\rVert \quad \text{ for } n \geq 1,\\
        \lVert \xi^{(n + 1)} - \xi^{(n)} \rVert & \leq \rho \lVert \xi^{(n)} - \xi^{(n - 1)}\rVert \quad \text{ for } n \geq 1,\\
        \lVert \xi - \xi^{(n)} \rVert & \leq \frac{\rho^n}{1 - \rho}\lVert\xi^{(1)}\rVert\,.
    \end{align}
\end{theorem}
This is essentially a variant of theorem V.2.11 in Ref.~\cite{stewart_sun_2004}. It remains to discuss the invertibility of $T$. We are thus interested in solving
\begin{align}
    \label{eq:pseudo_sylvester_equation}
    XL_2 - L_1\overline{X} = Y.
\end{align}
This closely resembles the Sylvester equation and the resolution can be done as follows. Write a Takagi factorization for $L_1$ and $L_2$: $L_1 = V\Lambda_1V^T$, $L_2 = W\Lambda_2W^T$ where $\Lambda_1 = \diag(\Lambda_{1i})$, $\Lambda_2 = \diag(\Lambda_{2i})$ are diagonal real positive. Then, a straightforward computation shows that Eq.~\eqref{eq:pseudo_sylvester_equation} is equivalent to
\begin{align}
    (V^\dagger XW)\Lambda_2 - \Lambda_1\overline{(V^\dagger XW)} = V^\dagger Y\overline{W}\,.
\end{align}
In case $\Lambda_1$ and $\Lambda_2$ have one common element (i.e $L_1$ and $L_2$ have one common singular value), say $\Lambda_{1i_0} = \Lambda_{2j_0}$, then substituting $(W^\dagger XV)_{ij} = \delta_{ii_0}\delta_{jj_0}$ in the above equation cancels the left-hand side, and then $T(X) = 0$ for the corresponding $X$. Thus, $T$ is singular.
On the contrary, if the singular values of $L_1$ and $L_2$ are pairwise distinct, there is the unique solution
\begin{align}
    (V^\dagger XW)_{ij} & = \frac{\Lambda_{2j}}{\Lambda_{2j}^2 - \Lambda_{1i}^2}(V^\dagger Y\overline{W})_{ij} + \frac{\Lambda_{1i}}{\Lambda_{2j}^2 - \Lambda_{1i}^2}\overline{(V^\dagger Y\overline{W})_{ij}} \,.
\end{align}
This shows that $\lVert T^{-1} \rVert$ is bounded (up to a constant) by $\inf_{i,j}\frac{1}{|\Lambda_{1i} - \Lambda_{2j}|}$.

Note that in case all the $\Lambda_{1i}$ are smaller than all the $\Lambda_{2i}$ (as it turns out to be the case with seesaw mechanisms),\footnote{Which implies that the singular values of $L_2$ are positive, hence $\overline{L_2}L_2 = L_2^\dagger L_2$ is nonsingular.} one can get an expression for $X$ as a power series without explicitly referring to the matrices $\Lambda, V, W$. To do this, expand $\frac{1}{\Lambda_{2j}^2 - \Lambda_{1i}^2} = \frac{1}{\Lambda_{2j}^2}\sum_{k \geq 0}\left(\frac{\Lambda_{1i}}{\Lambda_{2j}}\right)^{2k}$:
\begin{align}
    V^\dagger XW & = \sum_{k \geq 0}\left(\Lambda_1^{2k}V^\dagger Y\overline{W}\Lambda_2^{-2k - 1} + \Lambda_1^{2k + 1}V^T\overline{Y}W\Lambda_2^{-2k - 2}\right).
\end{align}
Then, substitute (from left to right) $\Lambda_1^{2k}V^\dagger \to V^\dagger(L_1\overline{L_1})^k$, $\overline{W}\Lambda_2^{-2k - 1} \to (\overline{L_2}L_2)^{-k}L_2^{-1}W$, $\Lambda_1^{2k + 1}V^T \to V^\dagger L_1(\overline{L_1}L_1)^k$, and $W\Lambda_2^{-2k - 2} \to (L_2\overline{L_2})^{-k - 1}W$. This yields
\begin{align}
    X & = \sum_{k \geq 0}(L_1\overline{L_1})^k(L_1\overline{Y} + Y\overline{L_2})(L_2\overline{L_2})^{-k - 1}\,,
\end{align}
and in particular $X = YL_2^{-1}$ in the special case where $L_1 = 0$.

Once Eq.~\eqref{eq:general_xi_definition} is solved, it remains to express $L_1'$ and $L_2'$ in terms of $L_1, L_2, G, \xi$. These expressions can take several forms:
\begin{align}
    L_2' & = (1 + \xi^\dagger\xi)^{-1/2}(L_2 + \xi^\dagger G + G^T\overline{\xi} + \xi^\dagger L_1\overline{\xi})(1 + \xi^T\overline{\xi})^{-1/2}\\
    & = (1 + \xi^\dagger\xi)^{1/2}(L_2 + G^T\overline{\xi})(1 + \xi^T\overline{\xi})^{-1/2}\\
    & = (1 + \xi^\dagger\xi)^{-1/2}(L_2 + \xi^\dagger G)(1 + \xi^T\overline{\xi})^{1/2}\\
    \label{eq:L2_nice_representation}
    & = L_2 + \xi^\dagger\frac{(1 + \xi\xi^\dagger)^{1/2} - 1}{\xi\xi^\dagger}G + G^T\frac{(1 + \overline{\xi}\xi^T)^{1/2} - 1}{\overline{\xi}\xi^T}\overline{\xi}\\
    &\quad + \xi^\dagger\frac{(1 + \xi\xi^\dagger)^{1/2} - \xi\xi^\dagger/2 - 1}{\xi\xi^\dagger}L_1\overline{\xi} + \xi^\dagger L_1\frac{(1 + \overline{\xi}\xi^T)^{1/2} - \overline{\xi}\xi^T/2 - 1}{\overline{\xi}\xi^T}\overline{\xi}\nonumber\\
    &\quad + \xi^\dagger\frac{(1 + \xi\xi^\dagger)^{1/2} - \xi\xi^\dagger - 1}{\xi\xi^\dagger}[(G + L_1\overline{\xi})\xi^T]\frac{(1 + \overline{\xi}\xi^T)^{-1/2} - 1}{\overline{\xi}\xi^T}\overline{\xi}\nonumber,\\
    L_1' & = (1 + \xi\xi^\dagger)^{-1/2}(L_1 - G\xi^T - \xi G^T + \xi L_2\xi^T)(1 + \overline{\xi}\xi^T)^{-1/2}\\
    & = (1 + \xi\xi^\dagger)^{-1/2}(L_1 - \xi G^T)(1 + \overline{\xi}\xi^T)^{1/2}\label{eq:L1_representation_2}\\
    & = (1 + \xi\xi^\dagger)^{1/2}(L_1 - G\xi^T)(1 + \overline{\xi}\xi^T)^{-1/2}.
\end{align}
When $L_1$ is simply a scalar, the last expression for $L_2'$ is particularly useful, since in this case $\xi\xi^\dagger$ is also a scalar. For the same reason, the last two expressions for $L_1'$ are interesting in this situation, since the non-integer powers cancel.

\subsection{Perturbation of singular vectors}
\label{sec:perturbation_singular_vectors}

In this section, we consider a symmetric matrix $A$ with singular vectors $x_i$ and associated singular values $\lambda_i$, i.e., $A\overline{x_i} = \lambda_ix_i$ for all $i$. Given a singular vector $x_i$, we want to know how this singular vector is perturbed when we add to $A$ a small perturbation $E$. We now fix $i$ once and for all; let then $X_i'$ contain (in columns) the $n_i - 1$ singular vectors $x_j$ such that $\lambda_i = \lambda_j$ and let $\hat{X}_i$ contain all the other $x_j$. (In other words, $n_i$ is the degeneracy of the singular value $\lambda_i$.) We will denote $X_i = \begin{pmatrix}
    x_i & X_i'
\end{pmatrix}$ and $X = \begin{pmatrix}
    X_i & \hat{X}_i
\end{pmatrix} = \begin{pmatrix}
    x_i & X_i' & \hat{X}_i
\end{pmatrix}$ and consider three cases:
\begin{itemize}
    \item The first one is when $\lambda_i$ is big (in a sense to be precised) and non-degenerate. The results are not explicitly used in the body of the article and are provided essentially for completeness and comparison with the second case.
    \item The second situation is when $\lambda_i$ is big but degenerate. This is useful to deal with the inverse seesaw.\footnote{We restrict ourselves to the case $\# \nu_L = \# N_R = \# S_R = 1$, so that $m_D$ and $\mu$ are scalar.} Indeed, when $\mu = 0$, one knows a simple Takagi factorization~\cite{Frere:1989xb} for the mass matrix of Eq.~\eqref{eq:ISmassmatrix}, but this is no longer the case when $\mu \neq 0$. For $\mu = 0$, the singular values of \eqref{eq:ISmassmatrix} (in other words, the physical masses) are $0, \sqrt{m_D^2 + M^2}, \sqrt{m_D^2 + M^2}$. The mass $\sqrt{m_D^2 + M^2}$ is thus twice degenerate but one can reasonably hope the degeneracy to be lifted when $\mu \neq 0$. We thus need to study how the $\sqrt{m_D^2 + M^2}$ degenerate mass subspace is perturbed when $\mu \neq 0$.
    \item The last case is when $\lambda_i = 0$ and not degenerate. Understanding it will allow us to figure out, for instance, how the massless eigenstate of the inverse seesaw is perturbed when $\mu$ goes from $0$ to a small finite value.
\end{itemize}

\subsubsection{Case of a large non-degenerate singular value}

In this special case $X_i'$ is empty. First, we rephase $x_i$: $x_i \to y_i = e^{\i\varphi/2}x_i$ where $e^{\i\varphi} \equiv \frac{\lambda_i + x_i^\dagger E\overline{x_i}}{|\lambda_i + x_i^\dagger E\overline{x_i}|}$. We then apply the general singular subspace perturbation theory to
\begin{align}
    M' & = \begin{pmatrix}
        y_i & \hat{X}_i
    \end{pmatrix}^\dagger(A + E)\overline{\begin{pmatrix}
        y_i & \hat{X}_i
    \end{pmatrix}} = \begin{pmatrix}
        |\lambda_i + x_i^\dagger E\overline{x_i}| & y_i^\dagger E\overline{\hat{X}_i}\\
        \hat{X}_i^\dagger E\overline{y_i} & \hat{X}_i^\dagger E\overline{\hat{X}_i}
    \end{pmatrix}
\end{align}
with the following partitioning:
\begin{align}
    L_1 & = |\lambda_i + x_i^\dagger E\overline{x_i}|\,,&
    L_2 & = \hat{\Lambda}_i + \hat{X}_i^\dagger E\overline{\hat{X}_i}\,,&
    G & = y_i^\dagger E\overline{\hat{X}_i}\,.
\end{align}
Since both $L_1$ and $L_2$ are nonzero in this case, it helps to separate $\xi$ into its real and imaginary parts to write down Eq.~\eqref{eq:general_xi_definition}. Explicitly, $\xi^{(1)}$ satisfies:
\begin{align}
    \Re(\xi^{(1)})\Re(L_2 - |\lambda_i + x_i^\dagger E\overline{x_i}|) - \Im(\xi^{(1)})\Im(L_2) & = \Re(y_i^\dagger E\overline{\hat{X}_i})\,,\\
    \Re(\xi^{(1)})\Im(L_2) + \Im(\xi^{(1)})\Re(L_2 + |\lambda_i + x_i^\dagger E\overline{x_i}|) & = \Im(y_i^\dagger E\overline{\hat{X}_i})\,,
\end{align}
which we can also write in matrix form:
\begin{align}
    \begin{pmatrix}
        \Re(\xi^{(1)}) & \Im(\xi^{(1)})
    \end{pmatrix}\begin{pmatrix}
        \Re(L_2 - |\lambda_i + x_i^\dagger E\overline{x_i}|) & -\Im(L_2)\\
        \Im(L_2) & \Re(L_2 + |\lambda_i + x_i^\dagger E\overline{x_i}|)
    \end{pmatrix} = \begin{pmatrix}
        \Re(y_i^\dagger E\overline{\hat{X}_i}) & \Im(y_i^\dagger E\overline{\hat{X}_i})
    \end{pmatrix}.
\end{align}

In all the following, we assume $\lVert E \rVert$ is negligible both compared to $\lambda_i$ and to the minimal spacing $m$ between $\lambda_i$ and the other singular values of $A$; we set $v \equiv \frac{\lVert E \rVert}{m} + \frac{\lVert E \rVert}{\lambda_i} \ll 1$. This will allow us to write,\footnote{Note that in case $\lambda_i = 0$ that will be considered later, it would of course be really uninteresting to require $\lVert E \rVert \ll \lambda_i$! Yet, the example formula that follows can still be written if we define $v$ to be $\frac{\lVert E \rVert}{m} + \frac{\lVert E \rVert}{\inf_{j \neq i}\lambda_j}$ instead. We will come back to this case in further details in the dedicated section.} for instance, that $(\hat{\Lambda}_i + |\lambda_i + x_i^\dagger E\overline{x_i}| + \mathcal{O}(\lVert E \rVert))^{-1} = (\hat{\Lambda}_i + \lambda_i)^{-1} + \frac{1}{\lVert E \rVert}\mathcal{O}(v^2)$. The $2 \times 2$ block matrix above may be explicitly inverted thanks to the formula
\begin{align}
    \label{eq:44_block_matrix_inverse}
    \begin{pmatrix}
        T_{11} & T_{12}\\
        T_{21} & T_{22}
    \end{pmatrix}^{-1} & = \begin{pmatrix}
        T_{11}^{-1}[1 + T_{12}(T_{22} - T_{21}T_{11}^{-1}T_{12})^{-1}T_{21}T_{11}^{-1}] & -T_{11}^{-1}T_{12}(T_{22} - T_{21}T_{11}^{-1}T_{12})^{-1})\\
        -(T_{22} - T_{21}T_{11}^{-1}T_{12})^{-1}T_{21}T_{11}^{-1} & (T_{22} - T_{21}T_{11}^{-1}T_{12})^{-1}
    \end{pmatrix},
\end{align}
which holds provided $T_{11}$ and $T_{22} - T_{21}T_{11}^{-1}T_{12}$ are nonsingular, which will turn out to be the case in our situation.\footnote{This follows from the fact that $\lVert E \rVert$ is negligible compared to the spacing of $\lambda_i$ with the other singular values of $A$.}
One can successively approximate:
\begin{align}
    T_{11}^{-1} & = (\hat{\Lambda}_i - \lambda_i)^{-1} + \frac{1}{\lVert E \rVert}\mathcal{O}(v^2),\\
    (T_{22} - T_{21}T_{11}^{-1}T_{12})^{-1} & = (\hat{\Lambda}_i + \lambda_i)^{-1} + \frac{1}{\lVert E \rVert}\mathcal{O}(v^2),\\
    -T_{11}^{-1}T_{12}(T_{22} - T_{21}T_{11}^{-1}T_{12})^{-1} & = \frac{1}{\lVert E \rVert}\mathcal{O}(v^2),\\
    -(T_{22} - T_{21}T_{11}^{-1}T_{12})^{-1}T_{21}T_{11}^{-1} & = \frac{1}{\lVert E \rVert}\mathcal{O}(v^2).
\end{align}
Hence,
\begin{align}
    \begin{pmatrix}
        T_{11} & T_{12}\\
        T_{21} & T_{22}
    \end{pmatrix}^{-1} & = \begin{pmatrix}
        (\hat{\Lambda}_i - \lambda_i)^{-1} & 0\\
        0 & (\hat{\Lambda}_i + \lambda_i)^{-1}
    \end{pmatrix} + \frac{1}{\lVert E \rVert}\mathcal{O}(v^2)\,,
\end{align}
and finally,
\begin{align}
    \Re(\xi^{(1)}) & = \Re(y_i^\dagger E\overline{\hat{X_i}})(\hat{\Lambda}_i - \lambda_i)^{-1} + \frac{\lVert \hat{X}_i^\dagger E\overline{x_i} \rVert}{\lVert E \rVert}\mathcal{O}(v^2)\nonumber\\
    & = \Re\left(\sqrt{\frac{\left|\lambda_i + x_i^\dagger E\overline{x_i}\right|}{\lambda_i + x_i^\dagger E\overline{x_i}}}x_i^\dagger E\overline{\hat{X_i}}\right)(\hat{\Lambda}_i - \lambda_i)^{-1} + \frac{\lVert \hat{X}_i^\dagger E\overline{x_i} \rVert}{\lVert E \rVert}\mathcal{O}(v^2)\\
    & = \frac{\lVert \hat{X}_i^\dagger E\overline{x_i} \rVert}{\lVert E \rVert}\mathcal{O}(v),\\
    \Im(\xi^{(1)}) & = \Im(y_i^\dagger E\overline{\hat{X_i}})(\hat{\Lambda}_i + \lambda_i)^{-1} + \frac{\lVert \hat{X}_i^\dagger E\overline{x_i} \rVert}{\lVert E \rVert}\mathcal{O}(v^2)\nonumber\\
    & = \Im\left(\sqrt{\frac{\left|\lambda_i + x_i^\dagger E\overline{x_i}\right|}{\lambda_i + x_i^\dagger E\overline{x_i}}}x_i^\dagger E\overline{\hat{X_i}}\right)(\hat{\Lambda}_i + \lambda_i)^{-1} + \frac{\lVert \hat{X}_i^\dagger E\overline{x_i} \rVert}{\lVert E \rVert}\mathcal{O}(v^2)\\
    & = \frac{\lVert \hat{X}_i^\dagger E\overline{x_i} \rVert}{\lVert E \rVert}\mathcal{O}(v).
\end{align}
Now, by the general perturbation theorem \ref{th:general_xi_equation_theorem},
\begin{align}
    \xi & = \xi^{(1)} + \left(\frac{\lVert \hat{X}_i^\dagger E\overline{y_i} \rVert}{\lVert E \rVert}\right)^3\mathcal{O}(v^3)\,,
\end{align}
and we can replace $\xi^{(1)}$ by the main term of our estimate in the formula above. It remains to determine how much the perturbed singular vector $\widetilde{x_i} = (e^{\i\varphi/2}x_i - \hat{X}_i\xi^\dagger)(1 + \xi\xi^\dagger)^{-1/2}$ computed from this $\xi$ must be rephased in order to make its singular value real positive. But from Eq.~\eqref{eq:L1_representation_2}, $L_1' = |\lambda_i + x_i^\dagger E\overline{x_i}| + \frac{\lVert \hat{X}_i^\dagger E\overline{y_i} \rVert^2}{\lVert E \rVert}\mathcal{O}(v)$ and since $v = \frac{\lVert E \rVert}{\lambda_i} \ll 1$, one obtains that the phase by which we need to rephase is $\left(\frac{\lVert\hat{X}_i^\dagger E\overline{y_i}\rVert}{\lVert E \rVert}\right)^2\mathcal{O}(v^2)$. Consequently, $e^{\i\varphi/2}x_i - \hat{X}_i(\xi^{(1)})^\dagger$ is, up to $\frac{\lVert\hat{X}_i^\dagger E\overline{y_i}\rVert}{\lVert E \rVert}\mathcal{O}(v^2)$, a perturbed singular vector with real positive singular value.

\subsubsection{Case of a zero non-degenerate singular value}
\label{sec:perturbation_zero_nondegenerate}
The general argument proceeds in the same lines as in the previous section. The first notable difference is with the definition of $v$,
\begin{align}
    v & = \frac{\lVert E \rVert}{m} + \frac{\lVert E \rVert}{\inf_{j \neq i}\lambda_j}\,,
\end{align}
instead of $v = \frac{\lVert E \rVert}{m} + \frac{\lVert E \rVert}{\lambda_i}$. Also, since $\lambda_i = 0$, $x_i$ remains a singular vector of $A$ with \textit{real} singular value $0$ when rephased. Hence, one may suppose that $x_i$ is already properly phased and forget about the $e^{\i\varphi}$.

Up to this definition change, the estimates one gets for $\xi^{(1)}$ are identical. Since $\lambda_i = 0$, we can actually write them in a more compact form:
\begin{align}
    \xi^{(1)} & = x_i^\dagger E\overline{\hat{X}_i}\hat{\Lambda}_i^{-1} + \frac{\lVert \hat{X}_i^\dagger E\overline{x_i} \rVert}{\lVert E \rVert}\mathcal{O}(v^2)\,.
\end{align}
The tricky point is actually with the final rephasing, i.e.~in the determination of the phase to add to $(x_i - \hat{X}_i\xi^\dagger)(1 + \xi\xi^\dagger)$ to make its associated singular value real positive. Again, we try to estimate $L_1'$ with the dedicated Eq.~\eqref{eq:L1_representation_2}:
\begin{align}
    L_1' & = |x_i^\dagger E\overline{x_i}| - x_i^\dagger E\overline{\hat{X}_i}\hat{\Lambda}_i^{-1}\hat{X}_i^\dagger E\overline{x_i} + \frac{\lVert \hat{X}_i^\dagger E\overline{x_i} \rVert^2}{\lVert E \rVert}\mathcal{O}(v^2).
\end{align}
In general, the main term of the estimate above will not be real positive. If one denotes by $e^{\i\chi}$ the phasis of the main term, then a perturbed singular vector with real positive singular value is given by:
\begin{align}
    & e^{\i\chi/2}e^{\i\left|\mathcal{O}\left(\frac{\lVert \hat{X}_i^\dagger E\overline{x_i} \rVert^2}{\lVert E \rVert\left||x_i^\dagger E\overline{x_i}| - x_i^\dagger E\overline{\hat{X}_i}\hat{\Lambda}_i^{-1}\hat{X}_i^\dagger E\overline{x_i}\right|}v^2\right)\right|}(x_i - \hat{X}_i\xi^\dagger)(1 + \xi\xi^\dagger)^{-1/2}\nonumber\\
    & = e^{\i\chi/2}e^{\i\left|\mathcal{O}\left(\frac{\lVert \hat{X}_i^\dagger E\overline{x_i} \rVert^2}{\lVert E \rVert\left||x_i^\dagger E\overline{x_i}| - x_i^\dagger E\overline{\hat{X}_i}\hat{\Lambda}_i^{-1}\hat{X}_i^\dagger E\overline{x_i}\right|}v^2\right)\right|}\left(x_i - \hat{X}_i\hat{\Lambda}_i^{-1}\hat{X}_i^T\overline{E}x_i + \frac{\lVert \hat{X}_i^\dagger E\overline{x_i} \rVert}{\lVert E \rVert}\mathcal{O}(v^2)\right)\nonumber\\
    &\quad \times \left(1 + \left(\frac{\lVert \hat{X}_i^\dagger E\overline{x_i} \rVert}{\lVert E \rVert}\right)^2|\mathcal{O}(v^2)|\right).
\end{align}
Note that generally speaking, there is no guarantee that $\mathcal{O}\left(\frac{\lVert E \rVert}{\left||x_i^\dagger E\overline{x_i}| - x_i^\dagger E\overline{\hat{X}_i}\hat{\Lambda}_i^{-1}\hat{X}_i^\dagger E\overline{x_i}\right|}v^2\right)$ is $\mathcal{O}(v^2)$, so that the main term of the last line $x_i - \hat{X}_i\hat{\Lambda}_i^{-1}\hat{X}_i^T\overline{E}x_i$ may not give a ``first order expansion'' of the perturbed singular vector. 

\subsubsection{Case of a large degenerate singular value}
\label{sec:perturbation_large_degenerate}

We now turn on to consider the case of a large degenerate singular value, i.e.~$\gg \lVert E \rVert$. The first step in this case is to factorize the $n_i \times n_i$ matrix $\begin{pmatrix}
    x_i & X_i'
\end{pmatrix}^\dagger(A + E)\overline{\begin{pmatrix}
    x_i & X_i'
\end{pmatrix}}$, that one may loosely call the restriction of $A + E$ to the degenerate subspace. This means finding a $n_i \times n_i$ unitary matrix $U$ such that the vectors $X_iU$ satisfy:
\begin{align}
    \left(X_iU\right)^\dagger(A + E)\overline{X_iU} = \begin{pmatrix}
    \lambda_i + \varepsilon_1 & 0\\
    0 & \lambda_iI_{n_i - 1} + \varepsilon'
\end{pmatrix},
\end{align}
where $\varepsilon'$ is a real diagonal matrix.\footnote{$\varepsilon_1$ and $\varepsilon'$ measures the shifting of the singular values between $\begin{pmatrix}
    x_i & X_i'
\end{pmatrix}^\dagger A\overline{\begin{pmatrix}
    x_i & X_i'
\end{pmatrix}} = \lambda_iI_{n_i}$ and $\begin{pmatrix}
    x_i & X_i'
\end{pmatrix}^\dagger (A + E)\overline{\begin{pmatrix}
    x_i & X_i'
\end{pmatrix}}$.} 
We will subsequently write $\varepsilon = \diag(\varepsilon_1, \varepsilon') = \diag(\varepsilon_j)$. The above equality implies
\begin{align}
    \label{eq:degenerate_subspace_diagonalization}
    U^\dagger X_i^\dagger(A + E)\overline{X_i}X_i^T(\overline{A} + \overline{E})X_iU = \begin{pmatrix}
        \lambda_i^2 + 2\lambda_i\varepsilon_1 + \varepsilon_1^2 & 0\\
        0 & \lambda_i^2I_{n_i - 1} + 2\lambda_i\varepsilon' + \varepsilon'^2
    \end{pmatrix},
\end{align}
hence,
\begin{align}
    \begin{pmatrix}
        2\lambda_i\varepsilon_1 + \varepsilon_1^2 & 0\\
        0 & 2\lambda_i\varepsilon' + \varepsilon'^2
    \end{pmatrix} & = U^\dagger X_i^\dagger(A + E)\overline{X_i}X_i^T(\overline{A} + \overline{E})X_iU - U^\dagger X_i^\dagger A\overline{X_i}X_i^T \overline{A}X_iU\nonumber\\
    & = \lambda_iU^\dagger X_i^T\overline{E}X_iU + \lambda_iU^\dagger X_i^\dagger E\overline{X_i}U + U^\dagger X_i^\dagger E\overline{X_i}X_i^T\overline{E}X_iU.
\end{align}
This shows that $\varepsilon_j = \mathcal{O}(\lVert E \rVert)$, not too surprisingly. In all that which follows, we assume $\varepsilon_1$ is distinct from all the diagonal elements of $\varepsilon' = \diag(\varepsilon_j)_{j \geq 2}$. We then apply the general theory of singular subspace perturbation to
\begin{align}
    M & = \begin{pmatrix}
        X_iU & \hat{X}_i
    \end{pmatrix}^\dagger(A + E)\overline{\begin{pmatrix}
    X_iU & \hat{X}_i
    \end{pmatrix}}\nonumber = \begin{pmatrix}
        y_i & Y_i' & \hat{X}_i
    \end{pmatrix}^\dagger(A + E)\overline{\begin{pmatrix}
        y_i & Y_i' & \hat{X}_i
    \end{pmatrix}}\nonumber\\
    & = \begin{pmatrix}
        \lambda_i + \varepsilon_1 & 0 & y_i^\dagger E\overline{\hat{X}_i}\\
        0 & \lambda_i + \varepsilon' & Y_i'^\dagger E\overline{\hat{X}_i}\\
        \hat{X}_i^\dagger E\overline{y_i} & \hat{X}_i^\dagger E\overline{Y_i'} & \hat{\Lambda}_i + \hat{X}_i^\dagger E\overline{\hat{X}_i}
    \end{pmatrix},
\end{align}
where the parameters of the perturbation problem are:
\begin{align}
    L_1 & = \lambda_i + \varepsilon_1, &
    L_2 & = \begin{pmatrix}
        \lambda_i + \varepsilon' & Y_i'^\dagger E\overline{\hat{X}_i}\\
        \hat{X}_i^\dagger E\overline{Y_i'} & \hat{\Lambda}_i + \hat{X}_i^\dagger E\overline{\hat{X}_i}
    \end{pmatrix},&
    G & = \begin{pmatrix}
        0 & y_i^\dagger E\overline{\hat{X}_i}
    \end{pmatrix}.
\end{align}
The relation defining $\xi^{(1)}$ reads:
\begin{align}
    \Re(\xi^{(1)})\Re(L_2 - \lambda_i - \varepsilon_1) - \Im(\xi^{(1)})\Im(L_2 - \lambda_i - \varepsilon_1) & = \Re(G) \,,\\
    \Re(\xi^{(1)})\Im(L_2 - \lambda_i - \varepsilon_1) + \Im(\xi^{(1)})\Re(L_2 + \lambda_i + \varepsilon_1) & = \Im(G) \,.
\end{align}
These two coupled equation can be written in matrix form:
\begin{align}
    &\begin{pmatrix}
        \Re(\xi^{(1)}) & \Im(\xi^{(1)})
    \end{pmatrix}\nonumber\\
    & \times \begin{pmatrix}
        \varepsilon' - \varepsilon_1 & \Re(Y_i'^\dagger E\overline{\hat{X}_i}) & 0 & \Im(Y_i'^\dagger E\overline{\hat{X}_i})\\
        \Re(\hat{X}_i^\dagger E\overline{Y_i'}) & \hat{\Lambda}_i - \lambda_i - \varepsilon_1 + \Re(\hat{X}_i^\dagger E\overline{\hat{X}_i}) & \Im(\hat{X}_i^\dagger E\overline{Y_i'}) & 0\\
        0 & -\Im(Y_i'^\dagger E\overline{\hat{X}_i}) & 2\lambda_i + \varepsilon_1 + \varepsilon' & \Re(Y_i'^\dagger E\overline{\hat{X}_i})\\
        -\Im(\hat{X}_i^\dagger E\overline{Y_i'}) & 0 & \Re(\hat{X}_i^\dagger E\overline{Y_i'}) & \hat{\Lambda}_i + \lambda_i + \varepsilon_1 + \Re(\hat{X}_i^\dagger E\overline{\hat{X}_i})
    \end{pmatrix}\nonumber\\
    & = \begin{pmatrix}
        0 & \Re(y_i^\dagger E\overline{\hat{X}_i}) & 0 & \Im(y_i^\dagger E\overline{\hat{X}_i})
    \end{pmatrix}.
\end{align}
We want to find an approximation for the inverse of the $4 \times 4$ block matrix above. To do this, we partition it into $4$ superblocks $\begin{pmatrix}
    T_{11} & T_{12}\\
    T_{21} & T_{22}
\end{pmatrix}$, where each superblock $T_{ij}$ comprises $2 \times 2$ blocks. We will then use Eq.~\eqref{eq:44_block_matrix_inverse} to calculate the inverse. We first compute estimates for $T_{11}^{-1}$ and $(T_{22} - T_{21}T_{11}^{-1}T_{12})^{-1}$; the estimates for all the blocks of $\begin{pmatrix}
    T_{11} & T_{12}\\
    T_{21} & T_{22}
\end{pmatrix}^{-1}$ will follow easily. Again, the inverses of $T_{11}$ and $T_{22} - T_{21}T_{11}^{-1}T_{12}$ can be found using the closed form Eq.~\eqref{eq:44_block_matrix_inverse}. To write down these estimates, we need to define $m$ the minimal spacing between $\lambda_i$ and the diagonal elements of $\hat{\Lambda}_i$ and $\mu$ the minimal spacing between $\varepsilon_1$ and the diagonal element of $\varepsilon'$. Keeping these definitions in mind, we introduce the following parameters:
\begin{align}
    u  = \frac{\lVert E \rVert}{\mu}\,,\qquad
    v  = \frac{\lVert E \rVert}{m} + \frac{\lVert E \rVert}{\lambda_i}\,.
\end{align}
Note that by definition, we should not expect $u$ to be small; on the contrary, since $\varepsilon_j = \mathcal{O}(\lVert E \rVert)$, one must expect $u \gtrsim 1$. However, we will require $v$ to be small, as well as $uv < 1$. Using this, we can state the following estimates for $T_{11}^{-1}$ and $(T_{22} - T_{21}T_{11}^{-1}T_{12})^{-1}$:
\begin{align}
    T_{11}^{-1} & = \begin{pmatrix}
        (\varepsilon' - \varepsilon_1)^{-1} & -(\varepsilon' - \varepsilon_1)^{-1}\Re(Y_i'^\dagger E\overline{\hat{X}_i})(\hat{\Lambda}_i - \lambda_i)^{-1}\\
        -(\hat{\Lambda}_i - \lambda_i)^{-1}\Re(\hat{X}_i^\dagger E\overline{Y_i'})(\varepsilon' - \varepsilon_1) & (\hat{\Lambda}_i - \lambda_i)^{-1}
    \end{pmatrix}\nonumber\\
    & \quad+ \frac{1}{\lVert E \rVert}\mathcal{O}\begin{pmatrix}
        u^2v & u^2v^2\\
        u^2v^2 & uv^2
    \end{pmatrix} = \frac{1}{\lVert E \rVert}\mathcal{O}\begin{pmatrix}
        u & uv\\
        uv & v
    \end{pmatrix},
\end{align}
\begin{align}
    (T_{22} - T_{21}T_{11}^{-1}T_{12})^{-1} & = \begin{pmatrix}
        (2\lambda_i)^{-1} & 0\\
        0 & (\hat{\Lambda}_i + \lambda_i)^{-1}
    \end{pmatrix} + \frac{1}{\lVert E \rVert}\mathcal{O}\begin{pmatrix}
        uv^2 & v^2\\
        v^2 & v^2
    \end{pmatrix} = \frac{1}{\lVert E \rVert}\mathcal{O}\begin{pmatrix}
        v & v^2\\
        v^2 & v
    \end{pmatrix},
\end{align}
from which follows:
\begin{align}
    -(T_{22} - T_{21}T_{11}^{-1}T_{12})^{-1}T_{21}T_{11}^{-1} & = \begin{pmatrix}
        0 & 0\\
        (\lambda_i + \hat{\Lambda}_i)^{-1}\Im(\hat{X}_i^\dagger E\overline{Y_i'})(\varepsilon' - \varepsilon_1)^{-1} & 0
    \end{pmatrix}\nonumber + \frac{1}{\lVert E \rVert}\mathcal{O}\begin{pmatrix}
        uv^2 & v^2\\
        u^2v^2 & uv^2
    \end{pmatrix}\\
    & = \frac{1}{\lVert E \rVert}\mathcal{O}\begin{pmatrix}
        uv^2 & v^2\\
        uv & uv^2
    \end{pmatrix},
		\end{align}
\begin{align}
    -T_{11}^{-1}T_{12}(T_{22} - T_{21}T_{11}^{-1}T_{12})^{-1} & = \begin{pmatrix}
        0 & -(\varepsilon' - \varepsilon_1)^{-1}\Im(Y_i'^\dagger E\overline{\hat{X}_i})(\lambda_i + \hat{\Lambda}_i)^{-1}\\
        0 & 0
    \end{pmatrix}\nonumber + \frac{1}{\lVert E \rVert}\mathcal{O}\begin{pmatrix}
        uv^2 & u^2v^2\\
        v^2 & uv^2
    \end{pmatrix}\\
    & = \frac{1}{\lVert E \rVert}\mathcal{O}\begin{pmatrix}
        uv^2 & uv\\
        v^2 & uv^2
    \end{pmatrix},
		\end{align}
\begin{align}
    T_{11}^{-1}[1 + T_{12}(T_{22} -T_{21}T_{11}^{-1}T_{12})^{-1}T_{21}T_{11}^{-1}] & = T_{11}^{-1} + \frac{1}{\lVert E \rVert}\mathcal{O}\begin{pmatrix}
        u^2v & u^2v^2\\
        u^2v^2 & u^2v^3
    \end{pmatrix}\nonumber\\
    &\hspace*{-5cm} = \begin{pmatrix}
        (\varepsilon' - \varepsilon_1)^{-1} & -(\varepsilon' - \varepsilon_1)^{-1}\Re(Y_i'^\dagger E\overline{\hat{X}_i})(\hat{\Lambda}_i - \lambda_i)^{-1}\\
        -(\hat{\Lambda}_i - \lambda_i)^{-1}\Re(\hat{X}_i^\dagger E\overline{Y_i'})(\varepsilon' - \varepsilon_1)^{-1} & (\hat{\Lambda}_i - \lambda_i)^{-1}
    \end{pmatrix}\nonumber\\
    &\hspace*{-5cm}\quad + \frac{1}{\lVert E \rVert}\mathcal{O}\begin{pmatrix}
        u^2v & u^2v^2\\
        u^2v^2 & uv^2
    \end{pmatrix}.
\end{align}
From all those estimates, one can finally find an approximation for $\xi^{(1)}$,
\begin{align}
\begin{split}
    \Re(\xi^{(1)})^T &= \begin{pmatrix}
        (\varepsilon' - \varepsilon_1)^{-1}\Re(Y_i'^\dagger E\overline{\hat{X}_i})(\lambda_i - \hat{\Lambda}_i)^{-1}\Re(\hat{X}_i^\dagger E\overline{y_i}) \\
        (\hat{\Lambda}_i - \lambda_i)^{-1}\Re(\hat{X}_i^\dagger E\overline{y_i})
    \end{pmatrix}\\
    & \quad+\begin{pmatrix}
         (\varepsilon' - \varepsilon_1)^{-1}\Im(Y_i'^\dagger E\overline{\hat{X}_i})(\lambda_i + \hat{\Lambda}_i)^{-1}\Im(\hat{X}_i^\dagger E\overline{y_i})\\
        0
    \end{pmatrix} + \frac{\lVert \hat{X}_i^\dagger E\overline{y_i} \rVert}{\lVert E \rVert}\mathcal{O}\begin{pmatrix}
        u^2v^2\\
        uv^2
    \end{pmatrix}\\
    & = \frac{\lVert \hat{X}_i^\dagger E\overline{y_i} \rVert}{\lVert E \rVert}\mathcal{O}\begin{pmatrix}
        uv\\
        v
    \end{pmatrix},\\
    \Im(\xi^{(1)})^T & = \begin{pmatrix}
        0\\
        (\hat{\Lambda}_i + \lambda_i)^{-1}\Im(\hat{X}_i^\dagger E\overline{y_i})
    \end{pmatrix} + \frac{\lVert \hat{X}_i^\dagger E\overline{y_i} \rVert}{\lVert E \rVert}\mathcal{O}\begin{pmatrix}
        uv^2\\
        v^2
    \end{pmatrix} = \frac{\lVert \hat{X}_i^\dagger E\overline{y_i} \rVert}{\lVert E \rVert}\begin{pmatrix}
        uv^2\\
        v
    \end{pmatrix}.
\end{split}
\label{eq:xi1approx}
\end{align}
By invoking theorem \ref{th:general_xi_equation_theorem}, one obtains
\begin{align}
    \xi & = \xi^{(1)} + \left(\frac{\lVert\hat{X}_i^\dagger E\overline{y_i}\rVert}{\lVert E \rVert}\right)^3\mathcal{O}(u^3v^2)= \xi^{(1)} + \left(\frac{\lVert\hat{X}_i^\dagger E\overline{y_i}\rVert}{\lVert E \rVert}\right)\mathcal{O}(u^3v^2)\,.
\end{align}
The $\mathcal{O}(u^3v^2)$ rest given by the general theorem is quite big, so that we can safely replace $\xi^{(1)}$ in the formula above by the estimate we computed in Eq.~\eqref{eq:xi1approx}. Then $\widetilde{x_i} = \left(y_i - \begin{pmatrix}Y_i' & \hat{X}_i\end{pmatrix}\xi^\dagger\right)(1 + \xi\xi^\dagger)^{-1/2}$ is a singular vector.

However, its associated singular value may not be real positive. Therefore, we still need to rephase it and to bound the required additional phase. Recalling Eq.~\eqref{eq:L1_representation_2}, one can estimate $L_1'$,
\begin{align}
    L_1' & = L_1 - G\xi^T  = \lambda_i + \varepsilon_1 + \frac{\lVert\hat{X}_i^\dagger E\overline{y_i}\rVert^2}{\lVert E \rVert}\mathcal{O}(v) = (\lambda_i + \varepsilon_1)\left(1 + \left(\frac{\lVert\hat{X}_i^\dagger E\overline{y_i}\rVert}{\lVert E \rVert}\right)^2\mathcal{O}(v^2)\right).
\end{align}
Consequently, our perturbed singular vector should be rephased by a phase not exceeding $\left(\frac{\lVert\hat{X}_i^\dagger E\overline{y_i}\rVert}{\lVert E \rVert}\right)^2\mathcal{O}(v^2)$, so that we have obtained a ``first-order'' expansion for the perturbed singular vector:
\begin{align}
    \widetilde{x_i} & = \left(y_i - \begin{pmatrix}
        Y_i' & \hat{X}_i
    \end{pmatrix}(\xi^{(1)})^\dagger + \frac{\lVert \hat{X}_i^\dagger E\overline{y_i}\rVert}{\lVert E \rVert}\mathcal{O}(u^3v^2)\right)  \left(1 + \left(\frac{\lVert \hat{X}_i^\dagger E\overline{y_i} \rVert}{\lVert E \rVert}\right)^2\mathcal{O}(u^2v^2)\right).
\end{align}
Note that the $\mathcal{O}$ in the last parentheses is scalar (it is a bound for $\xi\xi^\dagger$).

\subsection{Application to extended seesaw models}

Let us now apply our perturbation theory to our mass matrices of interest.

\subsubsection{Inverse seesaw}

The mass matrix of the inverse seesaw is given by Eq.~\eqref{eq:ISmassmatrix}. Since this is just a particular case of the extended inverse seesaw treated below, we will not go into the details but merely show the results. It is convenient to treat $\mu$ as a small perturbation, because the case $\mu=0$ can be factorized analytically~\cite{Frere:1989xb,Abada:2014vea}.
Following the procedure outlined above\footnote{More specifically, we use the results in section \ref{sec:perturbation_large_degenerate} to deal with the degenerate $\sqrt{m_D^2 + M^2}$ mass subspace and the results in section \ref{sec:perturbation_zero_nondegenerate} to figure out the perturbation of the massless eigenstate.} and assuming $M$, $m_D$, and $\mu$ to be real positive, we find
\begin{align}
\begin{pmatrix}
    0 & m_D & 0\\
    m_D & 0 & M\\
    0 & M & \mu
\end{pmatrix} \simeq U \,\text{diag}\left(\frac{\mu m_D^2}{M^2+m_D^2}, \sqrt{M^2+m_D^2}-\frac{\mu}{2}, \sqrt{M^2+m_D^2}+\frac{\mu}{2}\right) \,U^T\,,
\end{align}
up to terms of order $\mu m_D/M$ and $\mu^2/M$, with the mixing matrix
\begin{align}
U = \begin{pmatrix}
 \frac{M}{\sqrt{M^2+m_D^2}} & 
-\frac{\i m_D}{\sqrt{2}} \left(\frac{1}{\sqrt{M^2+m_D^2}}-\frac{5 M^2 \mu }{4 \left(M^2+m_D^2\right){}^2}\right) & 
\frac{m_D}{\sqrt{2}} \left(\frac{1}{\sqrt{M^2+m_D^2}}+\frac{5 M^2 \mu }{4 \left(M^2+m_D^2\right){}^2}\right) \\
 \frac{\mu  m_D}{M^2+m_D^2} & 
\frac{\i}{\sqrt{2}} \left(1+\frac{M^2 \mu }{4 \left(M^2+m_D^2\right){}^{3/2}}\right) & 
\frac{1}{\sqrt{2}}-\frac{M^2 \mu }{4 \sqrt{2}\left(M^2+m_D^2\right){}^{3/2}} \\
 -\frac{m_D}{\sqrt{M^2+m_D^2}} & 
-\frac{\i M}{\sqrt{2}} \left(\frac{1}{\sqrt{M^2+m_D^2}}-\frac{M^2 \mu }{4 \left(M^2+m_D^2\right){}^2}\right) & 
\frac{M}{ \sqrt{2}} \left(\frac{1}{\sqrt{M^2+m_D^2}}+\frac{M^2 \mu }{4 \left(M^2+m_D^2\right){}^2}\right) 
\end{pmatrix} .
\end{align}
Using this mixing matrix to rotate the majoron couplings gives back the result of Eq.~\eqref{eq:ISmajoron}.

\subsubsection{Extended inverse seesaw}
\label{sec:perturbation_eis}

In this part, we consider the case of the extended inverse seesaw, Eq.~\eqref{eq:EISmassmatrix}, where we suppose $M$ and $\mu_2$ to be real positive. Similar to the IS case above, one could simply perturb around $\mu_1 =\mu_2 = 0$~\cite{Abada:2014vea}. However, there are two potentially very different scales $\mu_1$ and $\mu_2$ at play in the perturbation, so the general perturbation theorem (which only takes into account the global magnitude of the perturbation) may not give satisfying estimates. We will therefore try a different method, forgetting about all the specific cases treated in section \ref{sec:perturbation_singular_vectors} and working only in the general framework of section \ref{sec:perturbation_singular_subspaces}. Essentially, we will apply the general theory of singular subspace perturbation to $M$ with the following partitioning:
\begin{align}
    L_1 & = 0\,,&
    L_2 & = \begin{pmatrix}
        \mu_1 & M\\
        M & \mu_2
    \end{pmatrix},&
    G & = \begin{pmatrix}
        m_D & 0
    \end{pmatrix},
\end{align}
i.e.~we perturb around a small $m_D$.
Since $L_1 = 0$, $\xi^{(1)}$ is simply given by $GL_2^{-1}$:
\begin{align}
    \xi^{(1)} & = \begin{pmatrix}
        -\frac{\mu_2m_D}{M^2 - \mu_1\mu_2} & \frac{m_DM}{M^2 - \mu_1\mu_2}
    \end{pmatrix}.
\end{align}
At this point it helps to define $M'^2 \equiv M^2 - \mu_1\mu_2$ and $m_D' \equiv m_D\frac{M^2}{M'^2}$ and to choose the phase of $m_D$ so that $m_D'$ be real positive. It will furthermore be convenient to define the small dimensionless parameters
\begin{align}
    \varepsilon = \frac{m_D'}{M}\,, && \eta_1 = \frac{|\mu_1|}{M}\,, && \eta_2 = \frac{\mu_2}{M}\,.
\end{align}
We want to bound the error between $\xi$ and $\xi^{(1)}$ component-wise (and not globally, contrary to what the general theorem \ref{th:general_xi_equation_theorem} permits). To achieve this, we first set
\begin{align}
    \label{eq:extended_iss_xin_components}
    \xi^{(n)} & \equiv \begin{pmatrix}
        \xi^{(n)}_1 & \xi^{(n)}_2
    \end{pmatrix} \equiv \begin{pmatrix}
        -\frac{\mu_2m_D'}{M^2}(1 + u_n) & \frac{m_D'}{M}(1 + v_n)
    \end{pmatrix}
\end{align}
for $n \geq 1$, where we expect $u_n, v_n$ to be small. By determining computationally the degrees (in $\varepsilon, \eta_1, \eta_2$) of the terms appearing in the first few $\xi^{(n)}$, we could conjecture some bounds on $u_n, v_n$. We will prove the following claim:
\begin{lemma}
    Assume $\varepsilon^2 < \frac{1}{6}, \eta_2^2 < \frac{1}{8}$. Then, for all $n \geq 1$, $|u_n| \leq 2\varepsilon^2$ and $|v_n| \leq 2\varepsilon^2\eta_2(\eta_1 + \eta_2)$.
    \begin{proof}
        Set $n \geq 1$ and write (component by component) the recursion equation giving $\xi^{(n + 1)}$ from $\xi^{(n)}$:
        \begin{align}
            \xi^{(n + 1)}_1 & = \xi^{(1)}_1 + m_D\xi^{(n)}_1\left(-\frac{\mu_2}{M'^2}\overline{\xi^{(n)}_1} + \frac{M}{M'^2}\overline{\xi^{(n)}_2}\right) ,\\
            \xi^{(n + 1)}_2 & = \xi^{(1)}_2 + m_D\xi^{(n)}_1\left(\frac{M}{M'^2}\overline{\xi^{(n)}_1} -\frac{\mu_1}{M'^2}\overline{\xi^{(n)}_2}\right) .
        \end{align}
        Then, expand the left-hand size using equation \ref{eq:extended_iss_xin_components}. This yields the following bounds on $u_{n + 1}$ and $v_{n + 1}$:
        \begin{align}
            |u_{n + 1}| & \leq (1 + |u_n|)\left(\varepsilon^2\eta_2^2(1 + |u_n|) + \varepsilon^2(1 + |v_n|)\right),\\
            |v_{n + 1}| & \leq (1 + |u_n|)\left(\varepsilon^2\eta_2^2(1 + |u_n|) + \varepsilon^2\eta_1\eta_2(1 + |v_n|)\right).
        \end{align}
        If one recalls $\varepsilon^2 < \frac{1}{6}$ and $\eta_2^2 < \frac{1}{8}$, the result follows immediately by induction.
    \end{proof}
\end{lemma}
This lemma finally allows us to write
\begin{align}
    \xi & = \begin{pmatrix}
        -\frac{\mu_2m_D'}{M^2} & \frac{m_D'}{M}
    \end{pmatrix} + \mathcal{O}\begin{pmatrix}
        \varepsilon^3\eta_2 & \varepsilon^3\eta_2(\eta_1 + \eta_2)
    \end{pmatrix}.
\end{align}
From this estimate, one can deduce approximations for $\xi\xi^\dagger$, $\xi^\dagger\xi$ and $L_2$ (for the last point, the easiest formula to use is Eq.~\eqref{eq:L2_nice_representation}).
\begin{align}
    \xi\xi^\dagger & = \frac{m_D'^2}{M^2}\left(1 + \frac{\mu_2^2}{M^2}\right) + \mathcal{O}(\varepsilon^4\eta_2(\eta_1 + \eta_2)),\\
    (1 + \xi\xi^\dagger)^{-1/2} & = \left(1 + \frac{m_D'^2}{M^2}\right)^{-1/2} + \mathcal{O}(\varepsilon^2\eta_2(\eta_2 + \varepsilon^2\eta_1)) = \mathcal{O}(\varepsilon^2),\\
    \xi^\dagger\xi & = \frac{m_D'^2}{M^2}\begin{pmatrix}
        \frac{\mu_2^2}{M^2} & -\frac{\mu_2}{M}\\
        -\frac{\mu_2}{M} & 1\\
    \end{pmatrix} + \mathcal{O}\begin{pmatrix}
        \varepsilon^4\eta_2^2 & \varepsilon^4\eta_2\\
        \varepsilon^4\eta_2 & \varepsilon^4\eta_2(\eta_1 + \eta_2)
    \end{pmatrix} = \mathcal{O}\begin{pmatrix}
        \varepsilon^2\eta_2^2 & \varepsilon^2\eta_2\\
        \varepsilon^2\eta_2 & \varepsilon^2
    \end{pmatrix},\\
    (1 + \xi^\dagger\xi)^{-1/2} & = \begin{pmatrix}
        1 - \frac{1}{2}\frac{m_D'^2\mu_2^2}{M^4} & \frac{1}{2}\frac{m_D'^2\mu_2}{M^3}\\
        \frac{1}{2}\frac{m_D'^2\mu_2}{M^3} & \left(1 + \frac{m_D'^2}{M^2}\right)^{-1/2}
    \end{pmatrix} + \mathcal{O}\begin{pmatrix}
        \varepsilon^4\eta_2^2 & \varepsilon^4\eta_2\\
        \varepsilon^4\eta_2 & \varepsilon^4\eta_2(\eta_1 + \eta_2)
    \end{pmatrix},\\
    \xi(1 + \xi^\dagger\xi)^{-1/2} & = \begin{pmatrix}
        -\frac{\mu_2 m_D'}{M^2} & \left(1 + \frac{m_D'^2}{M^2}\right)^{-1/2}\frac{m_D'}{M}
    \end{pmatrix} + \mathcal{O}\begin{pmatrix}
        \varepsilon^3\eta_2 & \varepsilon^3\eta_2(\eta_1 + \eta_2)
    \end{pmatrix},\\
    \xi^\dagger G + G^T\overline{\xi} & = \frac{m_D'}{M}\begin{pmatrix}
        -2\frac{\mu_2m_D}{M} & m_D\\
        m_D & 0
    \end{pmatrix} + M\mathcal{O}\begin{pmatrix}
        \varepsilon^4\eta_2 & \varepsilon^4\eta_2(\eta_1 + \eta_2)\\
        \varepsilon^4\eta_2(\eta_1 + \eta_2) & 0
    \end{pmatrix}\\
    & = M\mathcal{O}\begin{pmatrix}
        \varepsilon^2\eta_2 & \varepsilon^2\\
        \varepsilon^2 & 0
    \end{pmatrix},\\
    G\xi^T & = M\mathcal{O}(\varepsilon^2\eta_2),\\
    L_2' & = \begin{pmatrix}
        \left(1 + \frac{m_D'^2\mu_2^2}{M^4}\right)\mu_1 - \frac{m_D'^2}{M^2}\mu_2 & \sqrt{m_D'^2 + M^2} - \frac{1}{2}\frac{m_D'^2\mu_2}{M^3}\mu_1\\
        \sqrt{m_D'^2 + M^2} - \frac{1}{2}\frac{m_D'^2\mu_2}{M^3}\mu_1 & \mu_2
    \end{pmatrix}\\
    & \quad+ M\mathcal{O}\begin{pmatrix}
        \varepsilon^4\eta_2 & \varepsilon^4\eta_2(\eta_1 + \eta_2)\\
        \varepsilon^4\eta_2(\eta_1 + \eta_2) & \varepsilon^4\eta_2
    \end{pmatrix},\\
    L_1' & = \mu_2\frac{m_D'm_D}{M^2} + M\mathcal{O}(\varepsilon^4\eta_2).
\end{align}
The unitary matrix
\begin{align}
    U = \begin{pmatrix}
        (1 + \xi\xi^\dagger)^{-1/2} & \xi(1 + \xi^\dagger\xi)^{-1/2}\\
        -\xi^\dagger(1 + \xi\xi^\dagger)^{-1/2} & (1 + \xi^\dagger\xi)^{-1/2}
    \end{pmatrix}\begin{pmatrix}
        u_l & 0\\
        0 & U_h
    \end{pmatrix}
\end{align}
completely factorizes our mass matrix, where $u_l$ is the phase of $L_1'$ (which is $\sim 1$) and $U_h$ is the unitary matrix factorizing $L_2'$, which can be given in a closed form by Eq.~\eqref{eq:takagi_22_closed_form}. The final expression for the mixing matrix is not particularly illuminating, let us instead focus on the majoron couplings of interest, given by
\begin{align}
 -\frac{\i J}{2 f}\overline{N_i}\left[\gamma_5\Re(\mu_2U_{3i}U_{3j} - \overline{\mu_1}U_{2i}U_{2j}) + \i\Im(\mu_2U_{3i}U_{3j} - \overline{\mu_1}U_{2i}U_{2j})\right]N_j \equiv -\frac{J}{2f}\overline{N_i} B_{ij} N_j.
\end{align}
In the limit $\frac{m_D^2}{M^2} \ll \left|1 - \frac{|\mu_1|}{\mu_2}\right|$, the leading terms of the symmetric matrix $B$ are given by:
\begin{align}
\begin{pmatrix}
    \i\gamma_5\mu_2\frac{m_D'^2}{M^2} & -\frac{\mu_2m_D'}{\sqrt{2}M}\left(\i\gamma_5\Re\sqrt{\frac{\overline{\mu_1} + \mu_2}{|\overline{\mu_1} + \mu_2|}} - \Im\sqrt{\frac{\overline{\mu_1} + \mu_2}{|\overline{\mu_1} + \mu_2|}}\right) & -\frac{\mu_2m_D'}{\sqrt{2}M}\left(\i\gamma_5\Im\sqrt{\frac{\overline{\mu_1} + \mu_2}{|\overline{\mu_1} + \mu_2|}} + \Re\sqrt{\frac{\overline{\mu_1} + \mu_2}{|\overline{\mu_1} + \mu_2|}}\right)\\
    . & \i\gamma_5\frac{\mu_2^2 - |\mu_1|^2}{2|\overline{\mu_1} + \mu_2|} & -\i\gamma_5\frac{\mu_2\Im\mu_1}{|\overline{\mu_1} + \mu_2|} + \frac{1}{2}|\overline{\mu_1} + \mu_2|\\
    . & . & -\i\gamma_5\frac{\mu_2^2 - |\mu_1|^2}{2|\overline{\mu_1} + \mu_2|}\\
\end{pmatrix} .
\label{eq:EISmajoron_general}
\end{align}
Problems arise with our estimates only in case $|\mu_1| \sim \mu_2$ (or more accurately when the condition $\left|1 - \frac{|\mu_1|}{\mu_2}\right| \gg \varepsilon^2$ is violated). In this case we found numerically that the couplings $B_{22,23,33}$ do not actually go to zero for $\mu_1\to \pm \mu_2$ but are rather given by $m_1/2 = \mu_2 m_D'^2/(2 M^2)$, tiny but non-zero.
For real $\mu_1$ we get back the majoron couplings of Eq.~\eqref{eq:EISmajoron} used in the main text.

\bibliographystyle{JHEP}
\bibliography{BIB}

\providecommand{\href}[2]{#2}\begingroup\raggedright\begin{thebibliography}{10}

\bibitem{Hall:2009bx}
L.~J. Hall, K.~Jedamzik, J.~March-Russell and S.~M. West, \emph{{Freeze-In
  Production of FIMP Dark Matter}},
  \href{https://doi.org/10.1007/JHEP03(2010)080}{\emph{JHEP} {\bfseries 03}
  (2010) 080}, [\href{https://arxiv.org/abs/0911.1120}{{\ttfamily 0911.1120}}].

\bibitem{Bernal:2017kxu}
N.~Bernal, M.~Heikinheimo, T.~Tenkanen, K.~Tuominen and V.~Vaskonen, \emph{{The
  Dawn of FIMP Dark Matter: A Review of Models and Constraints}},
  \href{https://doi.org/10.1142/S0217751X1730023X}{\emph{Int. J. Mod. Phys.}
  {\bfseries A32} (2017) 1730023},
  [\href{https://arxiv.org/abs/1706.07442}{{\ttfamily 1706.07442}}].

\bibitem{Essig:2013goa}
R.~Essig, E.~Kuflik, S.~D. McDermott, T.~Volansky and K.~M. Zurek,
  \emph{{Constraining Light Dark Matter with Diffuse X-Ray and Gamma-Ray
  Observations}}, \href{https://doi.org/10.1007/JHEP11(2013)193}{\emph{JHEP}
  {\bfseries 11} (2013) 193},
  [\href{https://arxiv.org/abs/1309.4091}{{\ttfamily 1309.4091}}].

\bibitem{Bulbul:2014sua}
E.~Bulbul, M.~Markevitch, A.~Foster, R.~K. Smith, M.~Loewenstein and S.~W.
  Randall, \emph{{Detection of An Unidentified Emission Line in the Stacked
  X-ray spectrum of Galaxy Clusters}},
  \href{https://doi.org/10.1088/0004-637X/789/1/13}{\emph{Astrophys. J.}
  {\bfseries 789} (2014) 13},
  [\href{https://arxiv.org/abs/1402.2301}{{\ttfamily 1402.2301}}].

\bibitem{Boyarsky:2014jta}
A.~Boyarsky, O.~Ruchayskiy, D.~Iakubovskyi and J.~Franse, \emph{{Unidentified
  Line in X-Ray Spectra of the Andromeda Galaxy and Perseus Galaxy Cluster}},
  \href{https://doi.org/10.1103/PhysRevLett.113.251301}{\emph{Phys. Rev. Lett.}
  {\bfseries 113} (2014) 251301},
  [\href{https://arxiv.org/abs/1402.4119}{{\ttfamily 1402.4119}}].

\bibitem{Boyarsky:2014ska}
A.~Boyarsky, J.~Franse, D.~Iakubovskyi and O.~Ruchayskiy, \emph{{Checking the
  Dark Matter Origin of a 3.53 keV Line with the Milky Way Center}},
  \href{https://doi.org/10.1103/PhysRevLett.115.161301}{\emph{Phys. Rev. Lett.}
  {\bfseries 115} (2015) 161301},
  [\href{https://arxiv.org/abs/1408.2503}{{\ttfamily 1408.2503}}].

\bibitem{Iakubovskyi:2015dna}
D.~Iakubovskyi, E.~Bulbul, A.~R. Foster, D.~Savchenko and V.~Sadova,
  \emph{{Testing the origin of ~3.55 keV line in individual galaxy clusters
  observed with XMM-Newton}},
  \href{https://arxiv.org/abs/1508.05186}{{\ttfamily 1508.05186}}.

\bibitem{Cappelluti:2017ywp}
N.~Cappelluti, E.~Bulbul, A.~Foster, P.~Natarajan, M.~C. Urry, M.~W. Bautz
  et~al., \emph{{Searching for the 3.5 keV Line in the Deep Fields with
  Chandra: the 10 Ms observations}},
  \href{https://doi.org/10.3847/1538-4357/aaaa68}{\emph{Astrophys. J.}
  {\bfseries 854} (2018) 179},
  [\href{https://arxiv.org/abs/1701.07932}{{\ttfamily 1701.07932}}].

\bibitem{Anderson:2014tza}
M.~E. Anderson, E.~Churazov and J.~N. Bregman, \emph{{Non-Detection of X-Ray
  Emission From Sterile Neutrinos in Stacked Galaxy Spectra}},
  \href{https://doi.org/10.1093/mnras/stv1559}{\emph{Mon. Not. Roy. Astron.
  Soc.} {\bfseries 452} (2015) 3905--3923},
  [\href{https://arxiv.org/abs/1408.4115}{{\ttfamily 1408.4115}}].

\bibitem{Malyshev:2014xqa}
D.~Malyshev, A.~Neronov and D.~Eckert, \emph{{Constraints on 3.55 keV line
  emission from stacked observations of dwarf spheroidal galaxies}},
  \href{https://doi.org/10.1103/PhysRevD.90.103506}{\emph{Phys. Rev.}
  {\bfseries D90} (2014) 103506},
  [\href{https://arxiv.org/abs/1408.3531}{{\ttfamily 1408.3531}}].

\bibitem{Sekiya:2015jsa}
N.~Sekiya, N.~Y. Yamasaki and K.~Mitsuda, \emph{{A Search for a keV Signature
  of Radiatively Decaying Dark Matter with Suzaku XIS Observations of the X-ray
  Diffuse Background}}, \href{https://doi.org/10.1093/pasj/psv081}{\emph{Publ.
  Astron. Soc. Jap.} (2015) },
  [\href{https://arxiv.org/abs/1504.02826}{{\ttfamily 1504.02826}}].

\bibitem{Adhikari:2016bei}
R.~Adhikari et~al., \emph{{A White Paper on keV Sterile Neutrino Dark Matter}},
  \href{https://doi.org/10.1088/1475-7516/2017/01/025}{\emph{JCAP} {\bfseries
  1701} (2017) 025}, [\href{https://arxiv.org/abs/1602.04816}{{\ttfamily
  1602.04816}}].

\bibitem{Abazajian:2017tcc}
K.~N. Abazajian, \emph{{Sterile neutrinos in cosmology}},
  \href{https://doi.org/10.1016/j.physrep.2017.10.003}{\emph{Phys. Rept.}
  {\bfseries 711-712} (2017) 1--28},
  [\href{https://arxiv.org/abs/1705.01837}{{\ttfamily 1705.01837}}].

\bibitem{Bode:2000gq}
P.~Bode, J.~P. Ostriker and N.~Turok, \emph{{Halo formation in warm dark matter
  models}}, \href{https://doi.org/10.1086/321541}{\emph{Astrophys. J.}
  {\bfseries 556} (2001) 93--107},
  [\href{https://arxiv.org/abs/astro-ph/0010389}{{\ttfamily
  astro-ph/0010389}}].

\bibitem{Bullock:2017xww}
J.~S. Bullock and M.~Boylan-Kolchin, \emph{{Small-Scale Challenges to the
  $\Lambda$CDM Paradigm}},
  \href{https://doi.org/10.1146/annurev-astro-091916-055313}{\emph{Ann. Rev.
  Astron. Astrophys.} {\bfseries 55} (2017) 343--387},
  [\href{https://arxiv.org/abs/1707.04256}{{\ttfamily 1707.04256}}].

\bibitem{Fattahi:2016nld}
A.~Fattahi, J.~F. Navarro, T.~Sawala, C.~S. Frenk, L.~V. Sales, K.~Oman et~al.,
  \emph{{The cold dark matter content of Galactic dwarf spheroidals: no cores,
  no failures, no problem}},
  \href{https://arxiv.org/abs/1607.06479}{{\ttfamily 1607.06479}}.

\bibitem{Viel:2005qj}
M.~Viel, J.~Lesgourgues, M.~G. Haehnelt, S.~Matarrese and A.~Riotto,
  \emph{{Constraining warm dark matter candidates including sterile neutrinos
  and light gravitinos with WMAP and the Lyman-alpha forest}},
  \href{https://doi.org/10.1103/PhysRevD.71.063534}{\emph{Phys. Rev.}
  {\bfseries D71} (2005) 063534},
  [\href{https://arxiv.org/abs/astro-ph/0501562}{{\ttfamily
  astro-ph/0501562}}].

\bibitem{Merle:2014xpa}
A.~Merle and A.~Schneider, \emph{{Production of Sterile Neutrino Dark Matter
  and the 3.5 keV line}},
  \href{https://doi.org/10.1016/j.physletb.2015.07.080}{\emph{Phys. Lett.}
  {\bfseries B749} (2015) 283--288},
  [\href{https://arxiv.org/abs/1409.6311}{{\ttfamily 1409.6311}}].

\bibitem{Schneider:2016uqi}
A.~Schneider, \emph{{Astrophysical constraints on resonantly produced sterile
  neutrino dark matter}},
  \href{https://doi.org/10.1088/1475-7516/2016/04/059}{\emph{JCAP} {\bfseries
  1604} (2016) 059}, [\href{https://arxiv.org/abs/1601.07553}{{\ttfamily
  1601.07553}}].

\bibitem{Merle:2013wta}
A.~Merle, V.~Niro and D.~Schmidt, \emph{{New Production Mechanism for keV
  Sterile Neutrino Dark Matter by Decays of Frozen-In Scalars}},
  \href{https://doi.org/10.1088/1475-7516/2014/03/028}{\emph{JCAP} {\bfseries
  1403} (2014) 028}, [\href{https://arxiv.org/abs/1306.3996}{{\ttfamily
  1306.3996}}].

\bibitem{Merle:2015oja}
A.~Merle and M.~Totzauer, \emph{{keV Sterile Neutrino Dark Matter from Singlet
  Scalar Decays: Basic Concepts and Subtle Features}},
  \href{https://doi.org/10.1088/1475-7516/2015/06/011}{\emph{JCAP} {\bfseries
  1506} (2015) 011}, [\href{https://arxiv.org/abs/1502.01011}{{\ttfamily
  1502.01011}}].

\bibitem{Konig:2016dzg}
J.~K{\"o}nig, A.~Merle and M.~Totzauer, \emph{{keV Sterile Neutrino Dark Matter
  from Singlet Scalar Decays: The Most General Case}},
  \href{https://doi.org/10.1088/1475-7516/2016/11/038}{\emph{JCAP} {\bfseries
  1611} (2016) 038}, [\href{https://arxiv.org/abs/1609.01289}{{\ttfamily
  1609.01289}}].

\bibitem{Heeck:2017xbu}
J.~Heeck and D.~Teresi, \emph{{Cold keV dark matter from decays and
  scatterings}}, \href{https://doi.org/10.1103/PhysRevD.96.035018}{\emph{Phys.
  Rev.} {\bfseries D96} (2017) 035018},
  [\href{https://arxiv.org/abs/1706.09909}{{\ttfamily 1706.09909}}].

\bibitem{Bae:2017tqn}
K.~J. Bae, A.~Kamada, S.~P. Liew and K.~Yanagi, \emph{{Colder Freeze-in Axinos
  Decaying into Photons}},
  \href{https://doi.org/10.1103/PhysRevD.97.055019}{\emph{Phys. Rev.}
  {\bfseries D97} (2018) 055019},
  [\href{https://arxiv.org/abs/1707.02077}{{\ttfamily 1707.02077}}].

\bibitem{Bae:2017dpt}
K.~J. Bae, A.~Kamada, S.~P. Liew and K.~Yanagi, \emph{{Light axinos from
  freeze-in: production processes, phase space distributions, and Ly-$\alpha$
  forest constraints}},
  \href{https://doi.org/10.1088/1475-7516/2018/01/054}{\emph{JCAP} {\bfseries
  1801} (2018) 054}, [\href{https://arxiv.org/abs/1707.06418}{{\ttfamily
  1707.06418}}].

\bibitem{Kolb:1990vq}
E.~W. Kolb and M.~S. Turner, \emph{{The Early Universe}}, {\emph{Front. Phys.}
  {\bfseries 69} (1990) 1--547}.

\bibitem{Petraki:2007gq}
K.~Petraki and A.~Kusenko, \emph{{Dark-matter sterile neutrinos in models with
  a gauge singlet in the Higgs sector}},
  \href{https://doi.org/10.1103/PhysRevD.77.065014}{\emph{Phys. Rev.}
  {\bfseries D77} (2008) 065014},
  [\href{https://arxiv.org/abs/0711.4646}{{\ttfamily 0711.4646}}].

\bibitem{Frigerio:2011in}
M.~Frigerio, T.~Hambye and E.~Masso, \emph{{Sub-GeV dark matter as
  pseudo-Goldstone from the seesaw scale}},
  \href{https://doi.org/10.1103/PhysRevX.1.021026}{\emph{Phys. Rev.} {\bfseries
  X1} (2011) 021026}, [\href{https://arxiv.org/abs/1107.4564}{{\ttfamily
  1107.4564}}].

\bibitem{McDonald:2001vt}
J.~McDonald, \emph{{Thermally generated gauge singlet scalars as
  selfinteracting dark matter}},
  \href{https://doi.org/10.1103/PhysRevLett.88.091304}{\emph{Phys. Rev. Lett.}
  {\bfseries 88} (2002) 091304},
  [\href{https://arxiv.org/abs/hep-ph/0106249}{{\ttfamily hep-ph/0106249}}].

\bibitem{Baur:2015jsy}
J.~Baur, N.~Palanque-Delabrouille, C.~Y{\`e}che, C.~Magneville and M.~Viel,
  \emph{{Lyman-alpha Forests cool Warm Dark Matter}},
  \href{https://doi.org/10.1088/1475-7516/2016/08/012}{\emph{JCAP} {\bfseries
  1608} (2016) 012}, [\href{https://arxiv.org/abs/1512.01981}{{\ttfamily
  1512.01981}}].

\bibitem{Yeche:2017upn}
C.~Y{\`e}che, N.~Palanque-Delabrouille, J.~Baur and H.~du~Mas~des Bourboux,
  \emph{{Constraints on neutrino masses from Lyman-alpha forest power spectrum
  with BOSS and XQ-100}},
  \href{https://doi.org/10.1088/1475-7516/2017/06/047}{\emph{JCAP} {\bfseries
  1706} (2017) 047}, [\href{https://arxiv.org/abs/1702.03314}{{\ttfamily
  1702.03314}}].

\bibitem{Baur:2017stq}
J.~Baur, N.~Palanque-Delabrouille, C.~Y{\`e}che, A.~Boyarsky, O.~Ruchayskiy,
  {\'E}.~Armengaud et~al., \emph{{Constraints from Ly-$\alpha$ forests on
  non-thermal dark matter including resonantly-produced sterile neutrinos}},
  \href{https://doi.org/10.1088/1475-7516/2017/12/013}{\emph{JCAP} {\bfseries
  1712} (2017) 013}, [\href{https://arxiv.org/abs/1706.03118}{{\ttfamily
  1706.03118}}].

\bibitem{Irsic:2017ixq}
V.~Ir{\v s}i{\v c} et~al., \emph{{New Constraints on the free-streaming of warm
  dark matter from intermediate and small scale Lyman-$\alpha$ forest data}},
  \href{https://doi.org/10.1103/PhysRevD.96.023522}{\emph{Phys. Rev.}
  {\bfseries D96} (2017) 023522},
  [\href{https://arxiv.org/abs/1702.01764}{{\ttfamily 1702.01764}}].

\bibitem{Colombi:1995ze}
S.~Colombi, S.~Dodelson and L.~M. Widrow, \emph{{Large scale structure tests of
  warm dark matter}}, \href{https://doi.org/10.1086/176788}{\emph{Astrophys.
  J.} {\bfseries 458} (1996) 1},
  [\href{https://arxiv.org/abs/astro-ph/9505029}{{\ttfamily
  astro-ph/9505029}}].

\bibitem{Shaposhnikov:2006xi}
M.~Shaposhnikov and I.~Tkachev, \emph{{The nuMSM, inflation, and dark matter}},
  \href{https://doi.org/10.1016/j.physletb.2006.06.063}{\emph{Phys. Lett.}
  {\bfseries B639} (2006) 414--417},
  [\href{https://arxiv.org/abs/hep-ph/0604236}{{\ttfamily hep-ph/0604236}}].

\bibitem{Bezrukov:2014nza}
F.~Bezrukov and D.~Gorbunov, \emph{{Relic Gravity Waves and 7 keV Dark Matter
  from a GeV scale inflaton}},
  \href{https://doi.org/10.1016/j.physletb.2014.07.060}{\emph{Phys. Lett.}
  {\bfseries B736} (2014) 494--498},
  [\href{https://arxiv.org/abs/1403.4638}{{\ttfamily 1403.4638}}].

\bibitem{Tremaine:1979we}
S.~Tremaine and J.~E. Gunn, \emph{{Dynamical Role of Light Neutral Leptons in
  Cosmology}}, \href{https://doi.org/10.1103/PhysRevLett.42.407}{\emph{Phys.
  Rev. Lett.} {\bfseries 42} (1979) 407--410}.

\bibitem{Boyarsky:2008ju}
A.~Boyarsky, O.~Ruchayskiy and D.~Iakubovskyi, \emph{{A Lower bound on the mass
  of Dark Matter particles}},
  \href{https://doi.org/10.1088/1475-7516/2009/03/005}{\emph{JCAP} {\bfseries
  0903} (2009) 005}, [\href{https://arxiv.org/abs/0808.3902}{{\ttfamily
  0808.3902}}].

\bibitem{Olive:2016xmw}
{\scshape Particle Data Group} collaboration, C.~Patrignani et~al.,
  \emph{{Review of Particle Physics}},
  \href{https://doi.org/10.1088/1674-1137/40/10/100001}{\emph{Chin. Phys.}
  {\bfseries C40} (2016) 100001}.

\bibitem{Marsh:2015xka}
D.~J.~E. Marsh, \emph{{Axion Cosmology}},
  \href{https://doi.org/10.1016/j.physrep.2016.06.005}{\emph{Phys. Rept.}
  {\bfseries 643} (2016) 1--79},
  [\href{https://arxiv.org/abs/1510.07633}{{\ttfamily 1510.07633}}].

\bibitem{Irsic:2017yje}
V.~Ir{\v s}i{\v c}, M.~Viel, M.~G. Haehnelt, J.~S. Bolton and G.~D. Becker,
  \emph{{First constraints on fuzzy dark matter from Lyman-$\alpha$ forest data
  and hydrodynamical simulations}},
  \href{https://doi.org/10.1103/PhysRevLett.119.031302}{\emph{Phys. Rev. Lett.}
  {\bfseries 119} (2017) 031302},
  [\href{https://arxiv.org/abs/1703.04683}{{\ttfamily 1703.04683}}].

\bibitem{Armengaud:2017nkf}
E.~Armengaud, N.~Palanque-Delabrouille, D.~J.~E. Marsh, J.~Baur and
  C.~Y{\`e}che, \emph{{Constraining the mass of light bosonic dark matter using
  SDSS Lyman-$\alpha$ forest}},
  \href{https://doi.org/10.1093/mnras/stx1870}{\emph{Mon. Not. Roy. Astron.
  Soc.} {\bfseries 471} (2017) 4606--4614},
  [\href{https://arxiv.org/abs/1703.09126}{{\ttfamily 1703.09126}}].

\bibitem{Kobayashi:2017jcf}
T.~Kobayashi, R.~Murgia, A.~De~Simone, V.~Ir{\v s}i{\v c} and M.~Viel,
  \emph{{Lyman-$\alpha$ constraints on ultralight scalar dark matter:
  Implications for the early and late universe}},
  \href{https://doi.org/10.1103/PhysRevD.96.123514}{\emph{Phys. Rev.}
  {\bfseries D96} (2017) 123514},
  [\href{https://arxiv.org/abs/1708.00015}{{\ttfamily 1708.00015}}].

\bibitem{Hu:2000ke}
W.~Hu, R.~Barkana and A.~Gruzinov, \emph{{Cold and fuzzy dark matter}},
  \href{https://doi.org/10.1103/PhysRevLett.85.1158}{\emph{Phys. Rev. Lett.}
  {\bfseries 85} (2000) 1158--1161},
  [\href{https://arxiv.org/abs/astro-ph/0003365}{{\ttfamily
  astro-ph/0003365}}].

\bibitem{Hui:2016ltb}
L.~Hui, J.~P. Ostriker, S.~Tremaine and E.~Witten, \emph{{Ultralight scalars as
  cosmological dark matter}},
  \href{https://doi.org/10.1103/PhysRevD.95.043541}{\emph{Phys. Rev.}
  {\bfseries D95} (2017) 043541},
  [\href{https://arxiv.org/abs/1610.08297}{{\ttfamily 1610.08297}}].

\bibitem{Murgia:2017lwo}
R.~Murgia, A.~Merle, M.~Viel, M.~Totzauer and A.~Schneider, \emph{{"Non-cold"
  dark matter at small scales: a general approach}},
  \href{https://doi.org/10.1088/1475-7516/2017/11/046}{\emph{JCAP} {\bfseries
  1711} (2017) 046}, [\href{https://arxiv.org/abs/1704.07838}{{\ttfamily
  1704.07838}}].

\bibitem{Bezrukov:2014qda}
F.~Bezrukov and D.~Gorbunov, \emph{{Applicability of approximations used in
  calculations of the spectrum of dark matter particles produced in particle
  decays}}, \href{https://doi.org/10.1103/PhysRevD.93.063502}{\emph{Phys. Rev.}
  {\bfseries D93} (2016) 063502},
  [\href{https://arxiv.org/abs/1412.1341}{{\ttfamily 1412.1341}}].

\bibitem{Blas:2011rf}
D.~Blas, J.~Lesgourgues and T.~Tram, \emph{{The Cosmic Linear Anisotropy
  Solving System (CLASS) II: Approximation schemes}},
  \href{https://doi.org/10.1088/1475-7516/2011/07/034}{\emph{JCAP} {\bfseries
  1107} (2011) 034}, [\href{https://arxiv.org/abs/1104.2933}{{\ttfamily
  1104.2933}}].

\bibitem{Lesgourgues:2011rh}
J.~Lesgourgues and T.~Tram, \emph{{The Cosmic Linear Anisotropy Solving System
  (CLASS) IV: efficient implementation of non-cold relics}},
  \href{https://doi.org/10.1088/1475-7516/2011/09/032}{\emph{JCAP} {\bfseries
  1109} (2011) 032}, [\href{https://arxiv.org/abs/1104.2935}{{\ttfamily
  1104.2935}}].

\bibitem{Ade:2015xua}
{\scshape Planck} collaboration, P.~A.~R. Ade et~al., \emph{{Planck 2015
  results. XIII. Cosmological parameters}},
  \href{https://doi.org/10.1051/0004-6361/201525830}{\emph{Astron. Astrophys.}
  {\bfseries 594} (2016) A13},
  [\href{https://arxiv.org/abs/1502.01589}{{\ttfamily 1502.01589}}].

\bibitem{Rothstein:1992rh}
I.~Z. Rothstein, K.~S. Babu and D.~Seckel, \emph{{Planck scale symmetry
  breaking and majoron physics}},
  \href{https://doi.org/10.1016/0550-3213(93)90368-Y}{\emph{Nucl. Phys.}
  {\bfseries B403} (1993) 725--748},
  [\href{https://arxiv.org/abs/hep-ph/9301213}{{\ttfamily hep-ph/9301213}}].

\bibitem{Berezinsky:1993fm}
V.~Berezinsky and J.~W.~F. Valle, \emph{{The KeV majoron as a dark matter
  particle}}, \href{https://doi.org/10.1016/0370-2693(93)90140-D}{\emph{Phys.
  Lett.} {\bfseries B318} (1993) 360--366},
  [\href{https://arxiv.org/abs/hep-ph/9309214}{{\ttfamily hep-ph/9309214}}].

\bibitem{Lattanzi:2007ux}
M.~Lattanzi and J.~W.~F. Valle, \emph{{Decaying warm dark matter and neutrino
  masses}}, \href{https://doi.org/10.1103/PhysRevLett.99.121301}{\emph{Phys.
  Rev. Lett.} {\bfseries 99} (2007) 121301},
  [\href{https://arxiv.org/abs/0705.2406}{{\ttfamily 0705.2406}}].

\bibitem{Bazzocchi:2008fh}
F.~Bazzocchi, M.~Lattanzi, S.~Riemer-S{\o}rensen and J.~W.~F. Valle,
  \emph{{X-ray photons from late-decaying majoron dark matter}},
  \href{https://doi.org/10.1088/1475-7516/2008/08/013}{\emph{JCAP} {\bfseries
  0808} (2008) 013}, [\href{https://arxiv.org/abs/0805.2372}{{\ttfamily
  0805.2372}}].

\bibitem{Lattanzi:2013uza}
M.~Lattanzi, S.~Riemer-S{\o}rensen, M.~Tortola and J.~W.~F. Valle,
  \emph{{Updated CMB and x- and $\gamma$-ray constraints on Majoron dark
  matter}}, \href{https://doi.org/10.1103/PhysRevD.88.063528}{\emph{Phys. Rev.}
  {\bfseries D88} (2013) 063528},
  [\href{https://arxiv.org/abs/1303.4685}{{\ttfamily 1303.4685}}].

\bibitem{Queiroz:2014yna}
F.~S. Queiroz and K.~Sinha, \emph{{The Poker Face of the Majoron Dark Matter
  Model: LUX to keV Line}},
  \href{https://doi.org/10.1016/j.physletb.2014.06.016}{\emph{Phys. Lett.}
  {\bfseries B735} (2014) 69--74},
  [\href{https://arxiv.org/abs/1404.1400}{{\ttfamily 1404.1400}}].

\bibitem{Garcia-Cely:2017oco}
C.~Garcia-Cely and J.~Heeck, \emph{{Neutrino Lines from Majoron Dark Matter}},
  \href{https://doi.org/10.1007/JHEP05(2017)102}{\emph{JHEP} {\bfseries 05}
  (2017) 102}, [\href{https://arxiv.org/abs/1701.07209}{{\ttfamily
  1701.07209}}].

\bibitem{Chikashige:1980ui}
Y.~Chikashige, R.~N. Mohapatra and R.~D. Peccei, \emph{{Are There Real
  Goldstone Bosons Associated with Broken Lepton Number?}},
  \href{https://doi.org/10.1016/0370-2693(81)90011-3}{\emph{Phys. Lett.}
  {\bfseries B98} (1981) 265--268}.

\bibitem{Schechter:1981cv}
J.~Schechter and J.~W.~F. Valle, \emph{{Neutrino Decay and Spontaneous
  Violation of Lepton Number}},
  \href{https://doi.org/10.1103/PhysRevD.25.774}{\emph{Phys. Rev.} {\bfseries
  D25} (1982) 774}.

\bibitem{Wilczek:1982rv}
F.~Wilczek, \emph{{Axions and Family Symmetry Breaking}},
  \href{https://doi.org/10.1103/PhysRevLett.49.1549}{\emph{Phys. Rev. Lett.}
  {\bfseries 49} (1982) 1549--1552}.

\bibitem{Reiss:1982sq}
D.~B. Reiss, \emph{{Can the Family Group Be a Global Symmetry?}},
  \href{https://doi.org/10.1016/0370-2693(82)90647-5}{\emph{Phys. Lett.}
  {\bfseries B115} (1982) 217--220}.

\bibitem{Higaki:2014zua}
T.~Higaki, K.~S. Jeong and F.~Takahashi, \emph{{The 7 keV axion dark matter and
  the X-ray line signal}},
  \href{https://doi.org/10.1016/j.physletb.2014.04.007}{\emph{Phys. Lett.}
  {\bfseries B733} (2014) 25--31},
  [\href{https://arxiv.org/abs/1402.6965}{{\ttfamily 1402.6965}}].

\bibitem{Jaeckel:2014qea}
J.~Jaeckel, J.~Redondo and A.~Ringwald, \emph{{3.55 keV hint for decaying
  axionlike particle dark matter}},
  \href{https://doi.org/10.1103/PhysRevD.89.103511}{\emph{Phys. Rev.}
  {\bfseries D89} (2014) 103511},
  [\href{https://arxiv.org/abs/1402.7335}{{\ttfamily 1402.7335}}].

\bibitem{Kim:1986ax}
J.~E. Kim, \emph{{Light Pseudoscalars, Particle Physics and Cosmology}},
  \href{https://doi.org/10.1016/0370-1573(87)90017-2}{\emph{Phys. Rept.}
  {\bfseries 150} (1987) 1--177}.

\bibitem{Rojas:2017sih}
N.~Rojas, R.~A. Lineros and F.~Gonzalez-Canales, \emph{{Majoron dark matter
  from a spontaneous inverse seesaw model}},
  \href{https://arxiv.org/abs/1703.03416}{{\ttfamily 1703.03416}}.

\bibitem{Humbert:2015epa}
P.~Humbert, M.~Lindner and J.~Smirnov, \emph{{The Inverse Seesaw in Conformal
  Electro-Weak Symmetry Breaking and Phenomenological Consequences}},
  \href{https://doi.org/10.1007/JHEP06(2015)035}{\emph{JHEP} {\bfseries 06}
  (2015) 035}, [\href{https://arxiv.org/abs/1503.03066}{{\ttfamily
  1503.03066}}].

\bibitem{Mohapatra:1986aw}
R.~N. Mohapatra, \emph{{Mechanism for Understanding Small Neutrino Mass in
  Superstring Theories}},
  \href{https://doi.org/10.1103/PhysRevLett.56.561}{\emph{Phys. Rev. Lett.}
  {\bfseries 56} (1986) 561--563}.

\bibitem{Mohapatra:1986bd}
R.~N. Mohapatra and J.~W.~F. Valle, \emph{{Neutrino Mass and Baryon Number
  Nonconservation in Superstring Models}},
  \href{https://doi.org/10.1103/PhysRevD.34.1642}{\emph{Phys. Rev.} {\bfseries
  D34} (1986) 1642}.

\bibitem{GonzalezGarcia:1988rw}
M.~C. Gonzalez-Garcia and J.~W.~F. Valle, \emph{{Fast Decaying Neutrinos and
  Observable Flavor Violation in a New Class of Majoron Models}},
  \href{https://doi.org/10.1016/0370-2693(89)91131-3}{\emph{Phys. Lett.}
  {\bfseries B216} (1989) 360--366}.

\bibitem{Frere:1989xb}
J.~M. Fr{\`e}re and J.~Liu, \emph{{{CP} Violation in the Lepton Sector With
  Small Neutrino Masses}},
  \href{https://doi.org/10.1016/0550-3213(89)90469-0}{\emph{Nucl. Phys.}
  {\bfseries B324} (1989) 333--347}.

\bibitem{Abada:2014vea}
A.~Abada and M.~Lucente, \emph{{Looking for the minimal inverse seesaw
  realisation}},
  \href{https://doi.org/10.1016/j.nuclphysb.2014.06.003}{\emph{Nucl. Phys.}
  {\bfseries B885} (2014) 651--678},
  [\href{https://arxiv.org/abs/1401.1507}{{\ttfamily 1401.1507}}].

\bibitem{Barry:2011wb}
J.~Barry, W.~Rodejohann and H.~Zhang, \emph{{Light Sterile Neutrinos: Models
  and Phenomenology}},
  \href{https://doi.org/10.1007/JHEP07(2011)091}{\emph{JHEP} {\bfseries 07}
  (2011) 091}, [\href{https://arxiv.org/abs/1105.3911}{{\ttfamily 1105.3911}}].

\bibitem{Zhang:2011vh}
H.~Zhang, \emph{{Light Sterile Neutrino in the Minimal Extended Seesaw}},
  \href{https://doi.org/10.1016/j.physletb.2012.06.074}{\emph{Phys. Lett.}
  {\bfseries B714} (2012) 262--266},
  [\href{https://arxiv.org/abs/1110.6838}{{\ttfamily 1110.6838}}].

\bibitem{Heeck:2012bz}
J.~Heeck and H.~Zhang, \emph{{Exotic Charges, Multicomponent Dark Matter and
  Light Sterile Neutrinos}},
  \href{https://doi.org/10.1007/JHEP05(2013)164}{\emph{JHEP} {\bfseries 05}
  (2013) 164}, [\href{https://arxiv.org/abs/1211.0538}{{\ttfamily 1211.0538}}].

\bibitem{Ma:1995gf}
E.~Ma and P.~Roy, \emph{{Model of four light neutrinos prompted by all
  desiderata}}, \href{https://doi.org/10.1103/PhysRevD.52.R4780}{\emph{Phys.
  Rev.} {\bfseries D52} (1995) R4780--R4783},
  [\href{https://arxiv.org/abs/hep-ph/9504342}{{\ttfamily hep-ph/9504342}}].

\bibitem{Chun:1995js}
E.~J. Chun, A.~S. Joshipura and A.~{\relax Yu}. Smirnov, \emph{{Models of light
  singlet fermion and neutrino phenomenology}},
  \href{https://doi.org/10.1016/0370-2693(95)00967-P}{\emph{Phys. Lett.}
  {\bfseries B357} (1995) 608--615},
  [\href{https://arxiv.org/abs/hep-ph/9505275}{{\ttfamily hep-ph/9505275}}].

\bibitem{Dev:2012sg}
P.~S.~B. Dev and A.~Pilaftsis, \emph{{Minimal Radiative Neutrino Mass Mechanism
  for Inverse Seesaw Models}},
  \href{https://doi.org/10.1103/PhysRevD.86.113001}{\emph{Phys. Rev.}
  {\bfseries D86} (2012) 113001},
  [\href{https://arxiv.org/abs/1209.4051}{{\ttfamily 1209.4051}}].

\bibitem{Audren:2014bca}
B.~Audren, J.~Lesgourgues, G.~Mangano, P.~D. Serpico and T.~Tram,
  \emph{{Strongest model-independent bound on the lifetime of Dark Matter}},
  \href{https://doi.org/10.1088/1475-7516/2014/12/028}{\emph{JCAP} {\bfseries
  1412} (2014) 028}, [\href{https://arxiv.org/abs/1407.2418}{{\ttfamily
  1407.2418}}].

\bibitem{Poulin:2016nat}
V.~Poulin, P.~D. Serpico and J.~Lesgourgues, \emph{{A fresh look at linear
  cosmological constraints on a decaying dark matter component}},
  \href{https://doi.org/10.1088/1475-7516/2016/08/036}{\emph{JCAP} {\bfseries
  1608} (2016) 036}, [\href{https://arxiv.org/abs/1606.02073}{{\ttfamily
  1606.02073}}].

\bibitem{Gu:2009hn}
P.-H. Gu and U.~Sarkar, \emph{{Leptogenesis Bound on Spontaneous Symmetry
  Breaking of Global Lepton Number}},
  \href{https://doi.org/10.1140/epjc/s10052-011-1560-2}{\emph{Eur. Phys. J.}
  {\bfseries C71} (2011) 1560},
  [\href{https://arxiv.org/abs/0909.5468}{{\ttfamily 0909.5468}}].

\bibitem{Pilaftsis:1997jf}
A.~Pilaftsis, \emph{{CP violation and baryogenesis due to heavy Majorana
  neutrinos}}, \href{https://doi.org/10.1103/PhysRevD.56.5431}{\emph{Phys.
  Rev.} {\bfseries D56} (1997) 5431--5451},
  [\href{https://arxiv.org/abs/hep-ph/9707235}{{\ttfamily hep-ph/9707235}}].

\bibitem{Pilaftsis:2003gt}
A.~Pilaftsis and T.~E.~J. Underwood, \emph{{Resonant leptogenesis}},
  \href{https://doi.org/10.1016/j.nuclphysb.2004.05.029}{\emph{Nucl. Phys.}
  {\bfseries B692} (2004) 303--345},
  [\href{https://arxiv.org/abs/hep-ph/0309342}{{\ttfamily hep-ph/0309342}}].

\bibitem{Dev:2014laa}
P.~S. Bhupal~Dev, P.~Millington, A.~Pilaftsis and D.~Teresi, \emph{{Flavour
  Covariant Transport Equations: an Application to Resonant Leptogenesis}},
  \href{https://doi.org/10.1016/j.nuclphysb.2014.06.020}{\emph{Nucl. Phys.}
  {\bfseries B886} (2014) 569--664},
  [\href{https://arxiv.org/abs/1404.1003}{{\ttfamily 1404.1003}}].

\bibitem{Akhmedov:1998qx}
E.~K. Akhmedov, V.~A. Rubakov and A.~{\relax Yu}. Smirnov, \emph{{Baryogenesis
  via neutrino oscillations}},
  \href{https://doi.org/10.1103/PhysRevLett.81.1359}{\emph{Phys. Rev. Lett.}
  {\bfseries 81} (1998) 1359--1362},
  [\href{https://arxiv.org/abs/hep-ph/9803255}{{\ttfamily hep-ph/9803255}}].

\bibitem{Asaka:2005pn}
T.~Asaka and M.~Shaposhnikov, \emph{{The nuMSM, dark matter and baryon
  asymmetry of the universe}},
  \href{https://doi.org/10.1016/j.physletb.2005.06.020}{\emph{Phys. Lett.}
  {\bfseries B620} (2005) 17--26},
  [\href{https://arxiv.org/abs/hep-ph/0505013}{{\ttfamily hep-ph/0505013}}].

\bibitem{Hambye:2016sby}
T.~Hambye and D.~Teresi, \emph{{Higgs doublet decay as the origin of the baryon
  asymmetry}},
  \href{https://doi.org/10.1103/PhysRevLett.117.091801}{\emph{Phys. Rev. Lett.}
  {\bfseries 117} (2016) 091801},
  [\href{https://arxiv.org/abs/1606.00017}{{\ttfamily 1606.00017}}].

\bibitem{Haber:1984rc}
H.~E. Haber and G.~L. Kane, \emph{{The Search for Supersymmetry: Probing
  Physics Beyond the Standard Model}},
  \href{https://doi.org/10.1016/0370-1573(85)90051-1}{\emph{Phys. Rept.}
  {\bfseries 117} (1985) 75--263}.

\bibitem{Denner:1992vza}
A.~Denner, H.~Eck, O.~Hahn and J.~K{\"u}blbeck, \emph{{Feynman rules for
  fermion number violating interactions}},
  \href{https://doi.org/10.1016/0550-3213(92)90169-C}{\emph{Nucl. Phys.}
  {\bfseries B387} (1992) 467--481}.

\bibitem{stewart_sun_2004}
G.~Stewart and J.-G. Sun, \emph{Matrix Perturbation Theory}.
\newblock Academic Press, 1990.

\end{thebibliography}\endgroup

\end{document}